\documentclass[sigconf]{acmart}
\AtBeginDocument{%
  \providecommand\BibTeX{{%
    Bib\TeX}}}
\usepackage{xspace}
\usepackage{amsthm}
\usepackage{enumitem}
\usepackage{booktabs}
\usepackage{multirow}
\usepackage{graphicx}
\usepackage[normalem]{ulem}
\useunder{\uline}{\ul}{}
\usepackage{float}
\usepackage{subcaption}
\usepackage{balance}
\usepackage[toc,page]{appendix}

\newcommand{\modelname}{\textsf{SPMRec}\xspace}

\DeclareMathOperator*{\argmax}{arg\,max}
\DeclareMathOperator*{\argmin}{arg\,min}

\AtBeginDocument{%
  \providecommand\BibTeX{{%
    \normalfont B\kern-0.5em{\scshape i\kern-0.25em b}\kern-0.8em\TeX}}}
\begin{document}

%%
%% The "title" command has an optional parameter,
%% allowing the author to define a "short title" to be used in page headers.
% \title{Addressing the Rank Degeneration in Sequential Recommendation via Singular Spectrum Smoothing}
\title{Sequential Recommendation with Controllable Diversification: Representation Degeneration and Diversity}

%%
%% The "author" command and its associated commands are used to define
%% the authors and their affiliations.
%% Of note is the shared affiliation of the first two authors, and the
%% "authornote" and "authornotemark" commands
%% used to denote shared contribution to the research.
\author{Ziwei Fan}
% \email{zliu213@uic.edu}
\affiliation{%
  \institution{University of Illinois at Chicago}
  \country{USA}
}
\email{zfan20@uic.edu}

\author{Zhiwei Liu}
\affiliation{%
  \institution{Salesforce AI Research}
  \country{USA}
}
\email{zhiweiliu@salesforce.com}

\author{Hao Peng}
\affiliation{%
  \institution{Beihang University}
  \country{China}
}
\email{penghao@act.buaa.edu.cn}

\author{Philip S. Yu}
\affiliation{%
  \institution{University of Illinois at Chicago}
  \country{USA}
}
\email{psyu@uic.edu}

%%
%% By default, the full list of authors will be used in the page
%% headers. Often, this list is too long, and will overlap
%% other information printed in the page headers. This command allows
%% the author to define a more concise list
%% of authors' names for this purpose.
\renewcommand{\shortauthors}{Ziwei Fan, et al.}

%%
%% The abstract is a short summary of the work to be presented in the
%% article.
\begin{abstract}
  Sequential recommendation~(SR) models the dynamic user preferences and generates the next-item prediction as the affinity between the sequence and items, in a joint latent space with low dimensions (\textit{i.e.,} the sequence and item embedding space). Both sequence and item representations suffer from the representation degeneration issue due to the user\slash item long-tail distributions, where tail users\slash items are indistinguishably distributed as a narrow cone in the latent space. We argue that the representation degeneration issue is the root cause of insufficient recommendation diversity in existing SR methods, impairing the user potential exploration and further worsening the echo chamber issue.

  In this work, we first disclose the connection between the representation degeneration and recommendation diversity, in which severer representation degeneration indicates lower recommendation diversity. We then propose a novel \textbf{S}ingular s\textbf{P}ectrum s\textbf{M}oothing regularization for \textbf{Rec}ommendation~(\modelname), which acts as a controllable surrogate to alleviate the degeneration and achieve the balance between recommendation diversity and performance. The proposed smoothing regularization alleviates the degeneration by maximizing the area under the singular value curve, which is also the diversity surrogate. We conduct experiments on four benchmark datasets to demonstrate the superiority of \modelname, and show that the proposed singular spectrum smoothing can control the balance of recommendation performance and diversity simultaneously.
\end{abstract}

%%
%% The code below is generated by the tool at http://dl.acm.org/ccs.cfm.
%% Please copy and paste the code instead of the example below.
%%
% \begin{CCSXML}
% <ccs2012>
%  <concept>
%   <concept_id>10010520.10010553.10010562</concept_id>
%   <concept_desc>Computer systems organization~Embedded systems</concept_desc>
%   <concept_significance>500</concept_significance>
%  </concept>
%  <concept>
%   <concept_id>10010520.10010575.10010755</concept_id>
%   <concept_desc>Computer systems organization~Redundancy</concept_desc>
%   <concept_significance>300</concept_significance>
%  </concept>
%  <concept>
%   <concept_id>10010520.10010553.10010554</concept_id>
%   <concept_desc>Computer systems organization~Robotics</concept_desc>
%   <concept_significance>100</concept_significance>
%  </concept>
%  <concept>
%   <concept_id>10003033.10003083.10003095</concept_id>
%   <concept_desc>Networks~Network reliability</concept_desc>
%   <concept_significance>100</concept_significance>
%  </concept>
% </ccs2012>
% \end{CCSXML}

% \ccsdesc[500]{Computer systems organization~Embedded systems}
% \ccsdesc[300]{Computer systems organization~Redundancy}
% \ccsdesc{Computer systems organization~Robotics}
% \ccsdesc[100]{Networks~Network reliability}

%%
%% Keywords. The author(s) should pick words that accurately describe
%% the work being presented. Separate the keywords with commas.
\keywords{Recommendation, Nuclear Norm, Determinant, Diversity}

%% A "teaser" image appears between the author and affiliation
%% information and the body of the document, and typically spans the
%% page.

%%
%% This command processes the author and affiliation and title
%% information and builds the first part of the formatted document.
\maketitle

\begin{figure}
\begin{subfigure}[t]{0.235\textwidth}
    % \
    \includegraphics[width=\textwidth]{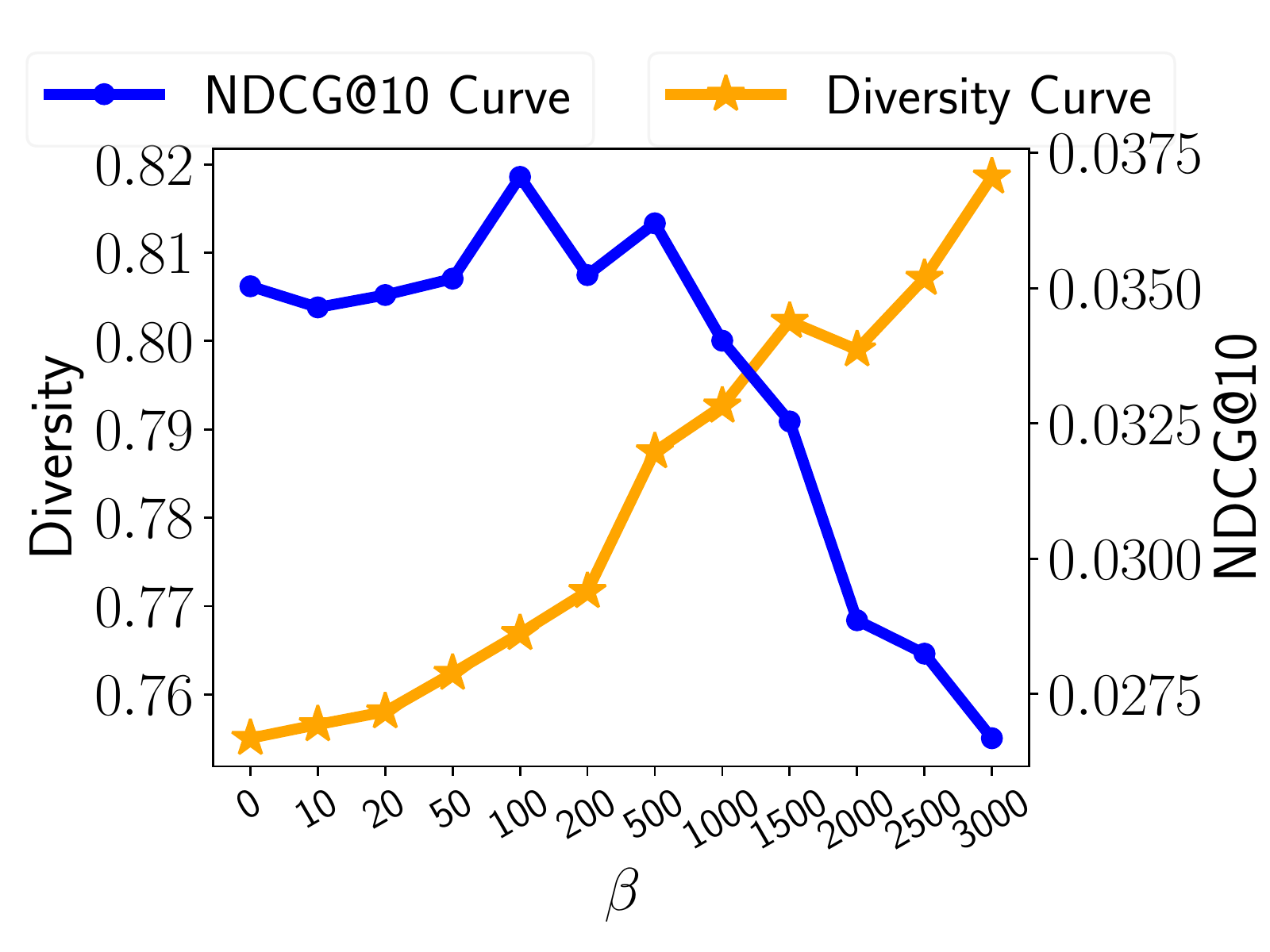}
    \caption{Beauty}
    \label{fig:beauty_item_diversity}
\end{subfigure}
\begin{subfigure}[t]{.235\textwidth}
    % \centering
    \includegraphics[width=\textwidth]{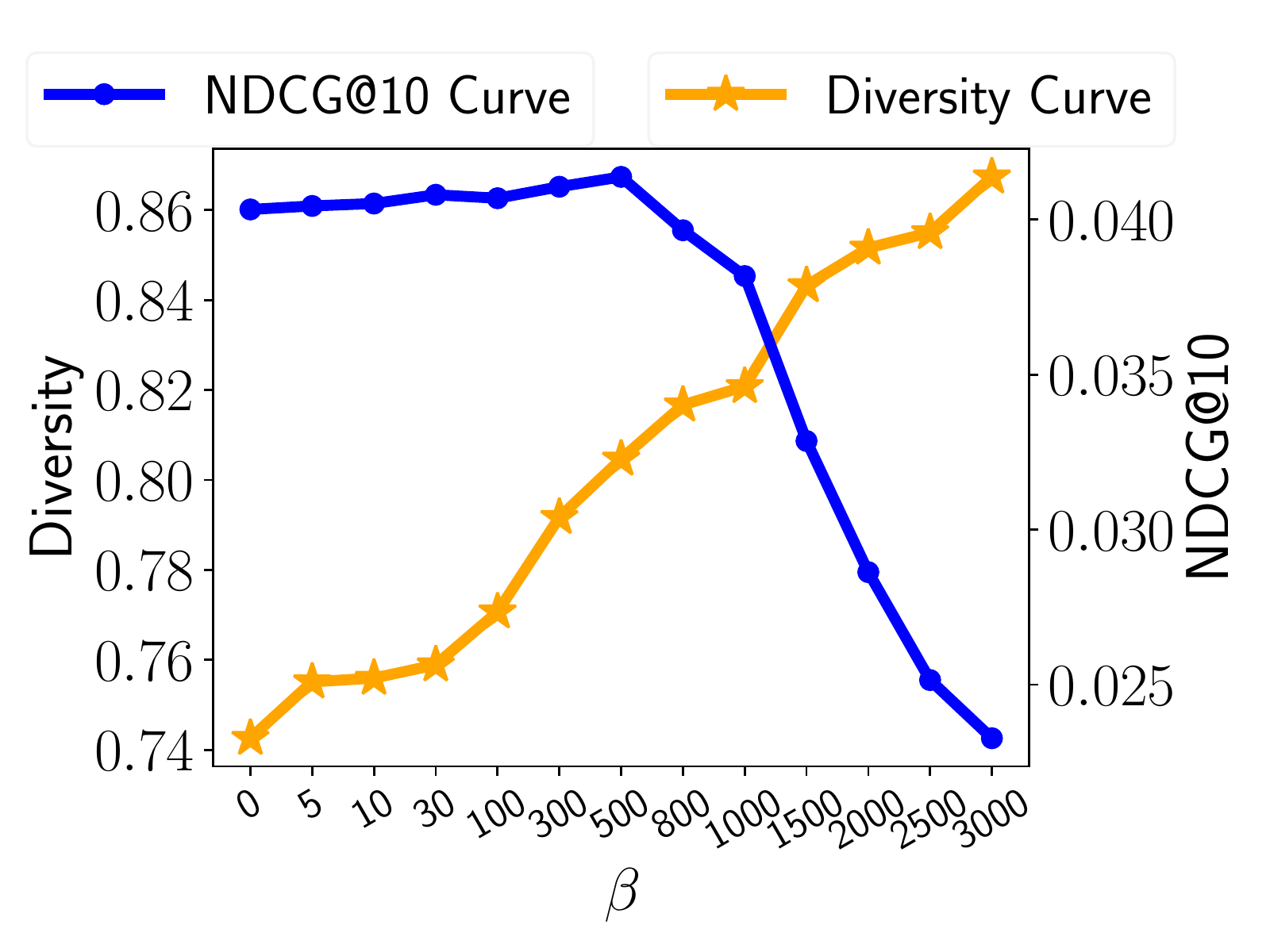}
    \caption{Toys}
    \label{fig:toys_item_diversity}
\end{subfigure}
% \vspace{-3mm}
\caption{A demonstrated strong correlation between the proposed spectrum smoothing regularization~(weighted by $\beta$) and intra-list diversity on public Amazon review data. A challenging balance between recommendation performance and diversity can be achieved with the proposed method.}
\label{fig:diversity_intro}
\end{figure}

\section{Introduction}
Recommender systems play a significant and widespread role in providing personalized recommendations and user modeling for various web applications, as evident in fashion, music, movies, and marketing scenarios.
Among existing frameworks, sequential recommendation~(SR)~\cite{rendle2010factorizing, he2016fusing, tang2018personalized, hidasi2015session, kang2018self} has recently garnered considerable attention due to its scalability and superior performance. SR utilizes sequential user behaviors and item-item transition correlations to predict the user's next preferred item. Among existing SR methods, Transformer-based methods~\cite{kang2018self, sun2019bert4rec, li2020time} have attracted increasing interests due to their capability of modeling high-order item-item transitions and scalable architecture design. 
In addition to recommendation quality (quantified as next-item prediction accuracy), a diverse set of items generation is also crucial for improving user satisfaction~\cite{zhang2008avoiding, chen2018fast} and helping users escape the echo chamber~\cite{ge2020understanding} for better long-term engagements~\cite{chen2021values}. 

The long-tail distribution of users\slash items has been widely observed in real-world applications and benchmarking datasets~\cite{zhang2023empowering, domingues2012combining, wang2022surrogate}, in which a small portion of users\slash items contribute to most interactions. According to empirical observations~\cite{goel2010anatomy}, tail items availability in recommendation can boost the sales of head items, which triggers the need of diverse recommendations. Demonstrated by~\cite{wang2018quantitative, 10.1145/3539618.3591636}, the long-tail distribution potentially leads to more intense Matthew effect, in which popular items and categories further dominate the recommender system with the implication of insufficient recommendation diversity. However, \textbf{few research investigates the relationship between the long-tail distribution of users\slash items and the recommendation diversity.}

Existing methods either focus on post re-ranking approaches~(with the risk of sacrificing recommendation quality)~\cite{zhang2008avoiding, chen2018fast, wilhelm2018practical} or build upon the indirect connection between recommendation quality and diversity~\cite{yang2023dgrec, liu2022determinantal}. A balance between recommendation quality and diversity is needed because over-emphasizing either the quality or the diversity leads to sub-optimal user satisfaction. To achieve this balance, \textbf{a recommendation model with controllable diversification module is needed}. It is also challenging to develop the controllable diversity module because the underlying connection between the quality and the diversity is not identified yet, and a differentiable solution is preferred so as to optimize both the quality and diversity.

\begin{figure}
\begin{subfigure}[t]{0.23\textwidth}
    % \
    \includegraphics[width=\textwidth]{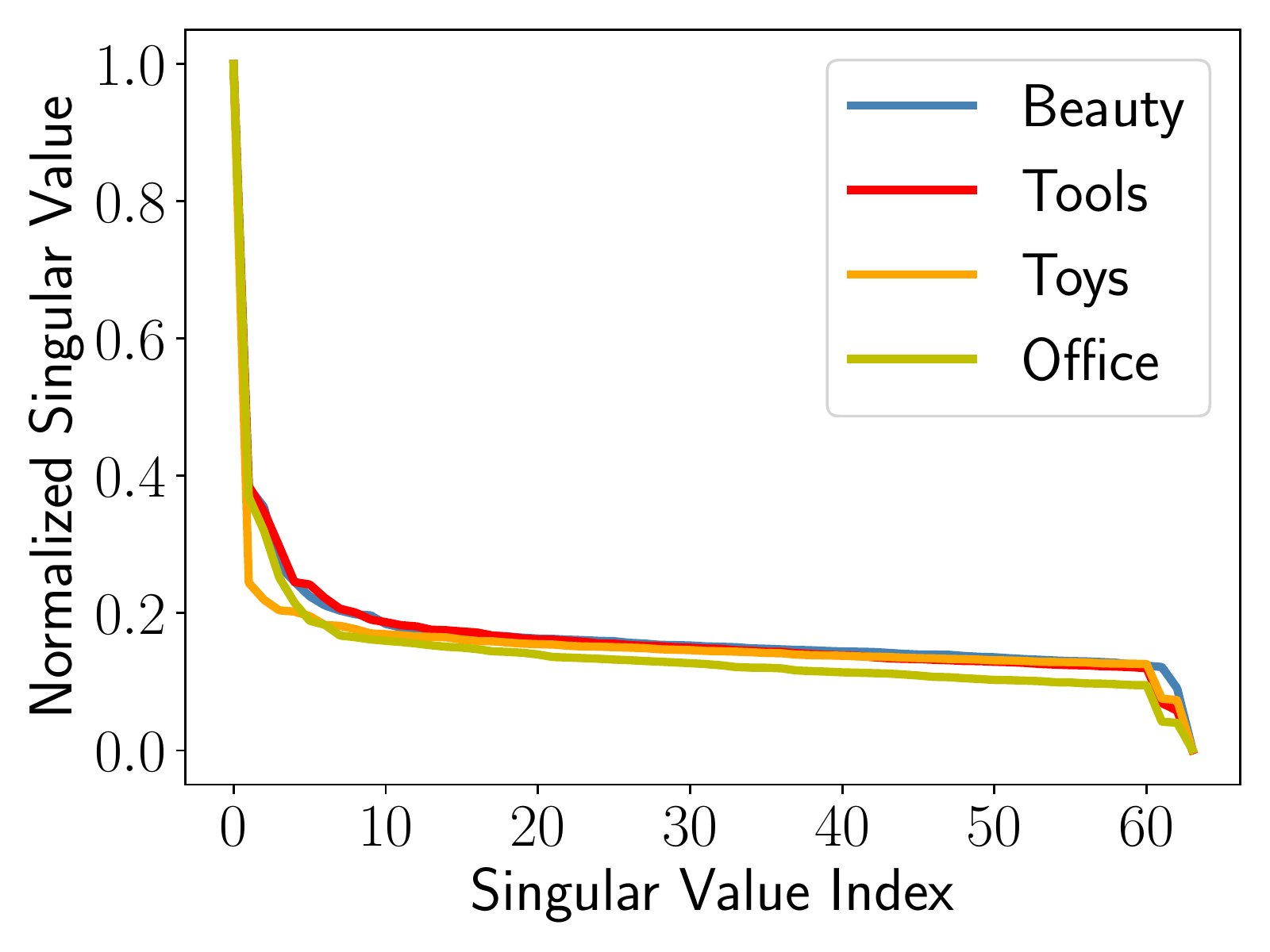}
    \caption{User Sequence Embeddings}
    \label{fig:beauty_user_moti_singular}
\end{subfigure}
% \hspace{8mm}
\begin{subfigure}[t]{.23\textwidth}
    % \centering
    \includegraphics[width=\textwidth]{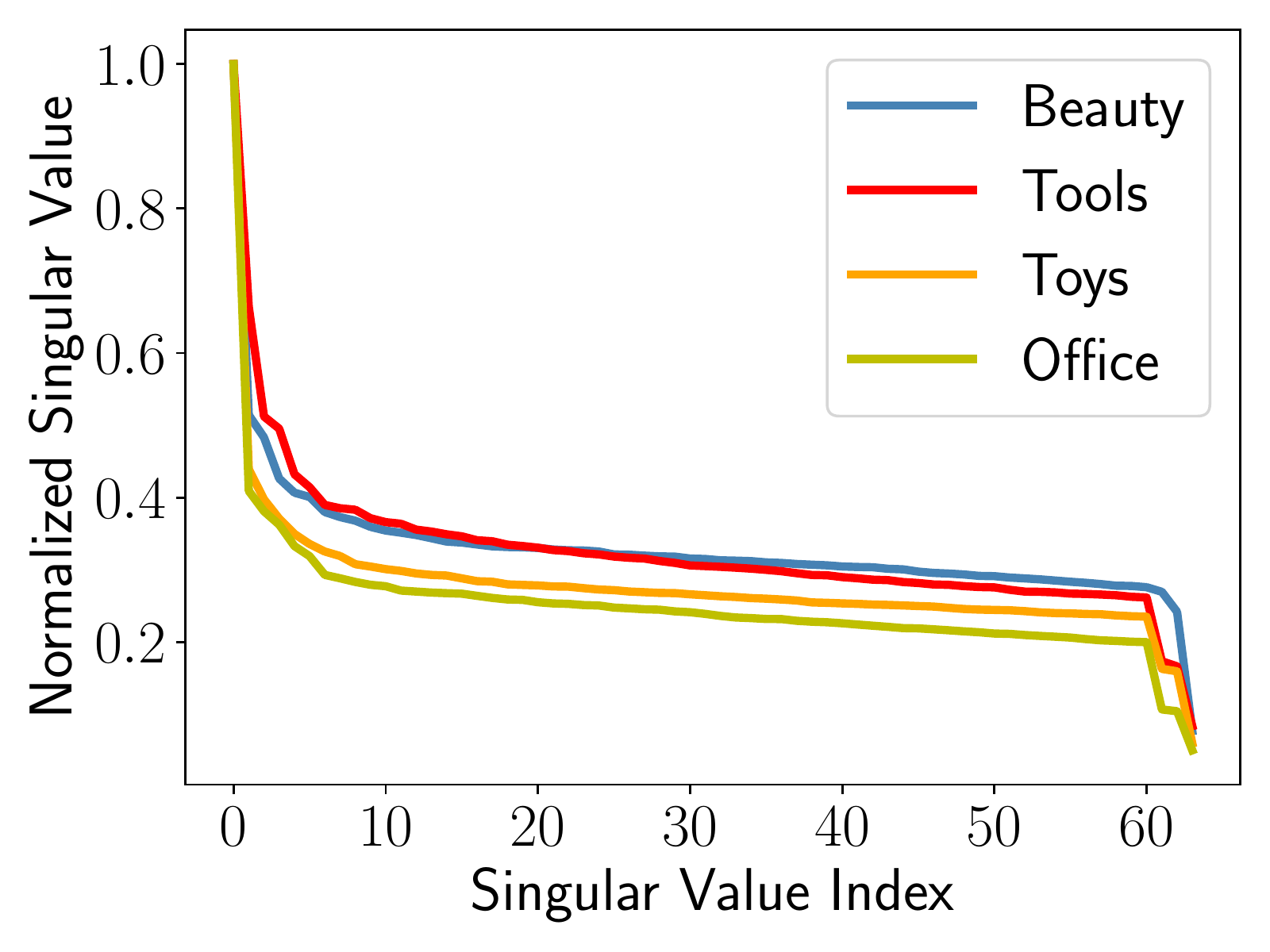}
    \caption{Item Embeddings}
    \label{fig:beauty_item_moti_singular}
\end{subfigure}
\caption{Fast singular values decay of user sequence output and item embeddings of SASRec are observed, implying the representation degeneration exists in both users and items. Singular values are normalized by dividing the largest singular value.}
\label{fig:motivation_singular}
\end{figure}

To this end, we demonstrate the connection between the long-tail distribution on users\slash items and the recommendation diversity, via the observed representation degeneration issue in both users and items, as shown in Fig.~(\ref{fig:motivation_singular}). We theoretically show that the fast singular value decay from the users\slash items representation degeneration is the root cause of insufficient diversity. More details will be explained in Section ~\ref{sec:degen_and_diverse}. Motivated by this connection, we develop a novel sequential recommendation with controllable diversity module by smoothing the singular value curves for users\slash items representations, with the goals of alleviating the degeneration and improving the recommendation diversity simultaneously.

Specifically, we propose a novel metric to measure the degree of degeneration issue and a novel \textbf{S}ingular s\textbf{P}ectrum s\textbf{M}oothing regularization design to achieve better sequence and item representation learning for \textbf{Rec}ommendations,~\textbf{\modelname}. We propose the \textbf{A}rea \textbf{U}nder the \textbf{S}ingular value \textbf{C}urve~(\textbf{AUSC}) to measure the degree of degeneration, with the intution that a flat curve has a larger area under the curve than a fast decaying curve, which will also be illustrated in the following Fig.~(\ref{fig:intuition_singular}). We further propose the singular spectrum smoothing regularization loss as an auxiliary loss to constrain the sequence and item embedding learning, which is a surrogate of our proposed AUSC metric. The singular spectrum smoothing regularization builds upon the nuclear norm and Frobenius norm~\cite{horn2012matrix, chen2022nuclear, DBLP:conf/cvpr/CuiWZLH020} to suppress the largest singular value and enlarge other singular values. The singular spectrum smoothing regularization unblocks the feasibility of sequential recommendation with controllable diversity.
We summarize our three technical contributions as follows:
\begin{itemize}[leftmargin=*]
    \item To the best of our knowledge, we are the first to identify the connection between the representation degeneration and insufficient diversity in SR from the theoretical perspective.
    \item We propose a novel metric~(AUSC) to measure the degree of degeneration and develop a singular spectrum smoothing regularization named \modelname to achieve the important but challenging balance of recommendation quality and diversity.
    \item Our experiments on four benchmark datasets show the effectiveness of the proposed \modelname, and empirically demonstrate that a proper contribution of smoothing improves both recommendation performance and diversity.
\end{itemize}

\section{Preliminaries}
% This section presents the formal definition of the SR problem and the adoption of the Transformer in SR. 
% The proposed singular spectrum smoothing \modelname applies to all Transformer-based SR methods.
\subsection{Problem Definition}
In the SR problem, there are a set of users $\mathcal{U}$, a set of items $\mathcal{V}$, and interactions feedback between users and items. Each interaction is associated with its timestamp. For each user $u\in\mathcal{U}$, we sort the interacted items chronologically and format the user as the sequence of interacted items, denoted as $\mathcal{S}^{u}={[v^{u}_1,v^u_2,\dots,v^u_{|\mathcal{S}^{u}|}]}$. Specifically, $v^{u}_i\in\mathcal{V}$ represents the $i$-th interacted item in the sequence $\mathcal{S}^{u}$ of the user $u$. SR tackles the problem of predicting the next item based on the sequence of user's historical interactions, which is formalized as predicting the probability of the next item $p\left( v_{|\mathcal{S}^{u}|+1}^{(u)}=v \left|  \mathcal{S}^{u} \right.\right)$. In the recommendation generation stage, the next item prediction scores over all items are ranked to generate the top-N recommendation list for the user $u$. 

\subsection{Sequential User Modeling}
\label{sec:transformer_basis}
% The fundamental challenge in SR is the sequential modeling of users' dynamic behaviors. Among existing sequential modeling methods for SR, Transformer-based methods~\cite{kang2018self,sun2019bert4rec} demonstrate superior effectiveness in capturing dynamic interests and modeling different orders of item-item transitions. Following the widely adopted Transformer architecture~\cite{vaswani2017attention}, Transformer-based SR methods consist of several essential components, including residual connection, layer normalization, self-attention module, and feed-forward networks. 

Given a user interaction sequence $\mathcal{S}^{u}$, we pre-process it by either truncating earliest interactions or padding with zeros such that the sequence length becomes the maximum sequence length $n$ to get a fixed length sequence $s=(s_1, s_2, \dots, s_n)$. An item embedding matrix $\mathbf{M}\in \mathbb{R}^{|\mathcal{V}|\times d}$ is randomly initialized and will be optimized, where $d$ is the latent dimension size. A trainable positional embedding $\mathbf{P}\in \mathbb{R}^{n\times d}$ is added to the sequentially ordered item embedding matrix as $\mathbf{E}_{\mathcal{S}^{u}}$. %:
% \begin{equation}
%     \label{eq:seq_emb}
%     \mathbf{E}_{\mathcal{S}^{u}} = [\mathbf{m}_{s_1}+\mathbf{p}_{s_1}, \mathbf{m}_{s_2}+\mathbf{p}_{s_2}, \dots, \mathbf{m}_{s_n}+\mathbf{p}_{s_n}].
% \end{equation}
% Following the original Transformer~\cite{vaswani2017attention}, existing Transformer-based sequential methods include \textit{Feed-forward Neural Networks}, \textit{Residual Connection}, and \textit{Layer Normalization} modules. 
Formally, we formulate the sequence encoder as follows:
\begin{align}
    \mathbf{H}_u = \text{SeqEnc}\left(\mathbf{E}_{\mathcal{S}^{u}}\right),
\end{align}
where $\mathbf{H}_u\in\mathbb{R}^{n\times d}$ denotes sequential output embeddings of $\mathcal{S}^u$, and for each timestep $t$, and $\textbf{H}_{(u,t)}$ encodes the predicted next-item~($v_{t+1}$) representation. 

\subsection{Next-Item Prediction Optimization}
The next item preference score prediction is given as:
\begin{align}
\label{eq:score_rec}
    p\left( v_{t+1}=v \left|  \mathcal{S}^{u} \right.\right) = \mathbf{H}_{(u,t)}^\top \mathbf{E}_v,
\end{align}
where $\mathbf{E}_v$ denotes the embedding of the item $v$.
The typical optimization process adopts the classical sampled cross-entropy~(sampled CE) loss or bayesian personalized ranking loss (BPR) \cite{rendle2012bpr}, which is maximizing the probability of users ranking positive items $v^{+}_{t+1}$ with greater preference scores over sampled negative items $v^{-}_{t+1}$. 
% Specifically, the maximization of probabilistic likelihood over all users and all time steps is as follows:
% \begin{align}
    % \text{max}\sum_{\mathcal{S}^u\in\mathcal{S}}\sum_{t=1}^{|\mathcal{S}^u|}p\left( v^{+}_{t+1} > v^{-}_{t+1}\left|  \mathcal{S}^{u} \right.\right),
% \end{align}
% where $v^{+}_{t+1} > v^{-}_{t+1}$ means that the positive item $v^{+}_{t+1}$ is ranked ahead of the negative item $v^{-}_{t+1}$,
The optimization can be formulated as a sampled cross-entropy loss minimization as:
\begin{align}
\label{eq:rec_loss}
    \mathcal{L}_{\text{rec}} =\sum_{\mathcal{S}^u\in\mathcal{S}}\sum_{t=1}^{|\mathcal{S}^u|}-\log\frac{\exp \left(\mathbf{H}_{(u,t)}^\top \mathbf{M}_{v^{+}_{t+1}}\right)}{\sum_{j=1}^{N^{-}_u}\exp\left(\mathbf{H}_{(u,t)}^\top \mathbf{M}_{v^{-}_{t+1}}^{j}\right)},
\end{align}
where $\mathbf{M}_{v^{-}_{t+1}}^{j}$ denotes the $j$-th negatively sampled item,  $N^{-}_u$ denotes the number of negatively sampled items from the item set without interaction with the user $u$. The value of $N^{-}_u$ affects the recommendation performance significantly~\cite{ding2019reinforced, chen2017sampling, mao2021simplex}. The typical choice in BPR and binary cross-entropy losses uses one negatively sampled item $N^{-}_u=1$. 

\section{Representation Degeneration and Diversity}
\label{sec:degen_and_diverse}
% In this section, we demonstrate the theoretical connection between the representation degeneration and the recommendation diversity. 

\subsection{Representation Degeneration Revisit}
\label{sec:degen_discussion}
\subsubsection{Revisit of Item Degeneration}
Most item embeddings are distributed in a narrow cone when the item degeneration issue occurs, with observation as fast singular value decay of the item embedding matrix $\mathbf{M}$~\cite{DBLP:conf/acl/YuSK0RY22, qiu2022contrastive}. 
As suggested in~\cite{DBLP:conf/iclr/GaoHTQWL19,DBLP:conf/acl/YuSK0RY22}, the long-tail distribution in frequency is the cause of representations degeneration. In SR, items follow the long-tail distribution. By utilizing the analysis in~\cite{DBLP:conf/iclr/GaoHTQWL19}, for any tail item $v$, the recommendation loss in Eq.~(\ref{eq:rec_loss}) associated with the learnable embedding $\mathbf{M}_v$ of the item $v$ is given as:
\begin{align}
    \min\sum_{\mathcal{S}^u\in\mathcal{S}}\sum_{t=1}^{|\mathcal{S}^u|}-\log\left(\exp \left(\mathbf{H}_{(u,t)}^\top \mathbf{M}_{v}\right)+C\right),
\end{align}
where $C$ includes the constant part without $\mathbf{M}_v$. The optimal solution is a vector with the uniformly negative direction of all sequence output embeddings $\mathbf{H}$ and infinite Frobenius norm, \textit{i.e.,} $||\mathbf{M}_v||\approx \infty$, using the Theorem 1 in~\cite{DBLP:conf/iclr/GaoHTQWL19}.
With the long-tail distribution of item frequency, most items are cold items. 
Most cold items degenerate into a narrow space as a cone, with the observation of fast singular decay phenomenon observed in~\cite{DBLP:conf/iclr/GaoHTQWL19, wang2020improving} and Fig.~(\ref{fig:beauty_item_moti_singular}). 
Moreover, as the Frobenius norm of item embedding matrix $||\mathbf{E}||$ is the upper bound of its largest singular value, the infinite Frobenius norm enlarges the largest singular value, causing the fast singular decay phenomenon observed in~\cite{DBLP:conf/iclr/GaoHTQWL19, wang2020improving}, which is also demonstrated in the Fig.~(\ref{fig:beauty_item_moti_singular}).

\subsubsection{Item Degeneration Induces Sequence Degeneration}
\label{sec:transformer_rankcollapse}
With degenerated item embeddings, we conclude that output embeddings of short sequences suffer from the degeneration issue. 
By freezing the item embedding matrix $\mathbf{M}$, we derive the recommendation loss of a sequence output embedding $\mathbf{H}_u$ for user $u$ as follows:
\begin{align}
    \min\sum_{t=1}^{|\mathcal{S}^u|}-\log\left(\exp \left(\mathbf{H}_{(u,t)}^\top \mathbf{M}_{v^{+}_{t+1}}\right)\right)+ \log\sum_{j=1}^{N^{-}_u}\exp\left(\mathbf{H}_{(u,t)}^\top \mathbf{M}_{v^{-}_{t+1}}^j\right),
\end{align}
where the first component enforces the $\mathbf{H}_u$ to be close to the positively interacted items' representations while the second part pulls the $\mathbf{H}_u$ away from negatively sampled items' representations. 

When the item degeneration happens, \textit{i.e.,} most tail items have $||\mathbf{M}_v||\approx \infty$, the optimal solution of $\mathbf{H}_u$ is a vector in the negative direction with most items, indicating $||\mathbf{H}_u||\approx 0$. This is because more negative items are sampled with more optimization steps. $||\mathbf{H}_u||\rightarrow 0$ indicates other singular values are being suppressed to 0, based on theories that the Frobenius norm is the upper bound of the largest singular value.
When $|\mathcal{S}^u|\approx 0$ (\textit{i.e.,} sequences are short), the degeneration issue becomes more severe, as there are much more negatively sampled items than positive items, where the second component dominates the learning process of $\mathbf{H}_u$.

\subsection{Theoretical Connection with Diversity}
\label{sec:theo_connect}

\subsubsection{Determinant of Embedding as Diversity Measurement}
We connect the determinant of embedding with the diversity via the well-defined determinant maximization problem in the determinant point process~(DPP)~\cite{kulesza2012determinantal}. 
The determinant maximization problem~\cite{kulesza2012determinantal} has been studied extensively in several machine learning applications, such as core-set discovery~\cite{mahabadi2019composable, mirrokni2015randomized}, feature selection~\cite{zadeh2017scalable}, and recommendation re-ranking~\cite{wilhelm2018practical, liu2022determinantal, chen2018fast, huang2021sliding}. Specifically, the determinant maximization problem is to find a subset from the total item set $\mathcal{Y}\in\mathcal{V}$ with $|\mathcal{Y}|\ll|\mathcal{V}|$. Formally, given the item feature matrix $\mathbf{F}\in\mathbb{R}^{|\mathcal{V}|\times d}$, DPP finds an item subset $\mathcal{Y}$, such that the determinant of the kernel matrix of the item subset $\text{det}(\mathbf{K}_{\mathcal{Y}})$ is maximized as follows:
\begin{align}
    \argmax_{\mathcal{Y}\in\mathcal{V}} \text{det}(\mathbf{K}_{\mathcal{Y}}),
\end{align}
where the kernel $\mathbf{K}_{\mathcal{Y}}(\cdot, \cdot)\in\mathbb{R}^{|\mathcal{Y}|\times|\mathcal{Y}|}$ is typically defined as the dot product of the item features $\mathbf{K}_{\mathcal{Y}}(\mathcal{Y}, \mathcal{Y})=\mathbf{F}_{\mathcal{Y}}^\top\mathbf{F}_{\mathcal{Y}}$. 
The existing DPP solutions~\cite{mahabadi2019composable, mirrokni2015randomized, wilhelm2018practical} iteratively expand the diverse subset by searching the item to enlarge the determinant of kernel $\text{det}(\mathbf{K}_{\mathcal{Y}})$. 

A subset with two items case (\textit{i.e.,} $\mathcal{Y}=\{v_i, v_j\}$) can be used to demonstrate the relationship between determinant of the kernel $\mathbf{K}_{\mathcal{Y}}$ and the diversity of the subset $\mathcal{Y}$. The determinant of the kernel $\text{det}(\mathbf{K}_{\mathcal{Y}})$ can be extended as $\text{det}(\mathbf{K}_{\mathcal{Y}})=\mathbf{K}_{v_iv_i}\mathbf{K}_{v_jv_j}-\mathbf{K}_{v_iv_j}\mathbf{K}_{v_jv_i}$, where $\mathbf{K}_{v_iv_j}\mathbf{K}_{v_jv_i}$ identifies the correlations with different items. When $\text{det}(\mathbf{K}_{\mathcal{Y}})$ obtains the minimum value 0, $\mathbf{K}_{v_iv_j}=\mathbf{K}_{v_iv_i}=\mathbf{K}_{v_jv_j}$, which indicates that $v_i$ and $v_j$ are highly correlated, \textit{i.e.,} no diversity. On the contrary, when $\text{det}(\mathbf{K}_{\mathcal{Y}})$ is maximized, the diverse set of items is retrieved. 

Extending to the total item set $\mathcal{V}$ for recommendations in SR, the total item set's diversity can be measured as the determinant of the kernel on $\mathcal{V}$ and the associated item embedding matrix $\mathbf{M}$, which is $\text{det}(\mathbf{K}_{\mathcal{V}}) = \text{det}(\mathbf{M}^\top\mathbf{M})$. Inductively, given a selected item set with the size as $|\mathcal{Y}|=n$, DPP adds an item $v_j$ to the selected item set $\mathcal{Y}$ as $\hat{\mathcal{Y}}$ with size $n+1$. Using Matrix Determinant Lemma, the the determinant of $\hat{\mathcal{Y}}$ is given as:
\begin{align}
    \text{det}(\hat{\mathcal{Y}})&=\text{det}\left(\begin{bmatrix}
    \mathbf{K}_{\mathcal{Y}} & \mathbf{K}_{\mathcal{Y}, v_j}=\mathbf{F}_{\mathcal{Y}}^\top\mathbf{F}_{v_j}\\
    \mathbf{K}_{v_j, \mathcal{Y}}=\mathbf{F}_{v_j}^\top\mathbf{F}_{\mathcal{Y}} & \mathbf{K}_{v_j, v_j}=\mathbf{F}_{v_j}^\top\mathbf{F}_{v_j}
    \end{bmatrix}\right)\\
    &= \text{det}(\mathbf{K}_{\mathcal{Y}})\cdot \text{det}\left(\mathbf{K}_{v_j, v_j} - \frac{\mathbf{K}_{v_j, \mathcal{Y}}\cdot \mathbf{K}_{\mathcal{Y}, v_j}}{\mathbf{K}_{\mathcal{Y}}}\right). 
\end{align}
$\mathbf{K}_{v_j, \mathcal{Y}}\cdot \mathbf{K}_{\mathcal{Y}, v_j}$ measures the correlations between the item $v_j$ and the selected item set $\mathcal{Y}$.
As $\mathcal{Y}$ is selected and fixed, $\text{det}(\mathbf{K}_{\mathcal{Y}})$ and $\mathbf{K}_{\mathcal{Y}}$ can be treated as constants, and $\mathbf{K}_{v_j, v_j}$ is the similarity of the item itself and can also be treated as a constant. DPP aims at finding the $v_j$ for maximizing the $\text{det}(\hat{\mathcal{Y}})$, which is equivalent to finding the $v_j$ with the least correlations with the selected item set $\mathcal{Y}$:
\begin{align}
    \argmax_{v_j} \text{det}(\mathbf{K}_{\hat{\mathcal{Y}}}) = \argmin_{v_j} \mathbf{K}_{v_j, \mathcal{Y}}\cdot \mathbf{K}_{\mathcal{Y}, v_j}.
\end{align}
This inductive argument implies that DPPs favor a diverse item set.

By applying the logorithmic and the SVD decomposition on the embedding matrix $\mathbf{M}$, we obtain the following the relationship between diversity and the singular values of $\mathbf{M}$:
\begin{align}
\label{eq:diverse_det}
    % \log\text{det}(\mathbf{K}_{\mathcal{V}}) = 
    \log\text{det}(\mathbf{M}^\top\mathbf{M}) = \log\prod_{i=1}^d \left(\sigma^{\mathbf{M}}_i\right)^2 = \sum_{i=1}^d2\log \sigma^{\mathbf{M}}_i \propto \sum_{i=1}^d\sigma^{\mathbf{M}}_i,
\end{align}
where $\left[\sigma^{\mathbf{M}}_1, \sigma^{\mathbf{M}}_2, \dots, \sigma^{\mathbf{M}}_d\right]$ are singular values of the item embedding matrix $\mathbf{M}$, and singular values are in the descending order as $\sigma^{\mathbf{M}}_1>\sigma^{\mathbf{M}}_2>\dots>\sigma^{\mathbf{M}}_d>0$ and the $\sigma^{\mathbf{M}}_1$ is the largest singular value $\sigma^{\mathbf{M}}_{\text{max}}$. With Eq.~(\ref{eq:diverse_det}), we establish the theoretical relationship between the determinant of embedding and item set diversity. As the 2log only re-scales the singular value, we have shown that the diversity is proportional to the sum of singular values, \textit{i.e.,} the nuclear norm of the embedding $\sum_{i=1}^d\sigma^{\mathbf{M}}_i=|\mathbf{M}|_{*}$.

\begin{figure}
    \includegraphics[width=0.48\textwidth]{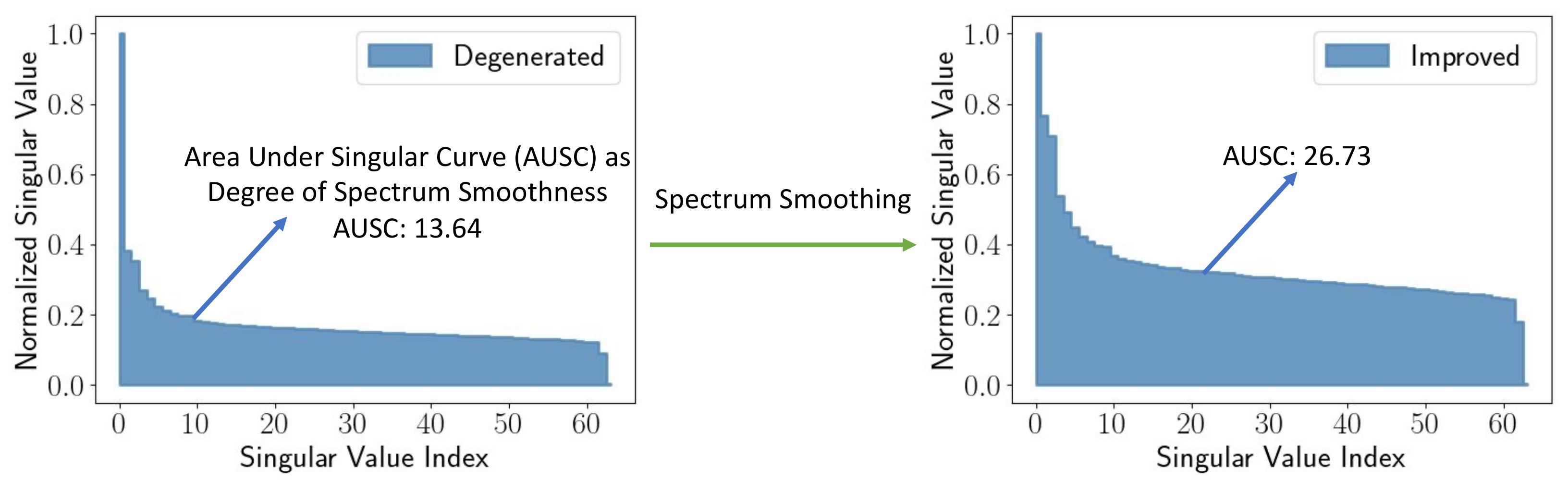}
    \caption{Intuition of Singular Spectrum Smoothing for addressing sequence and item representation degeneration.}
\label{fig:intuition_singular}
\end{figure}

\subsubsection{Representation Degeneration and Insufficient Diversity}
The observations in Fig.~(\ref{fig:motivation_singular}) show that both item embeddings and sequence output embeddings find the fast singular value decay, which is a crucial indicator of the representation degeneration issue~\cite{wang2020improving}. With SVD to decompose both sequence output embeddings $\mathbf{H}$ and item embeddings $\mathbf{M}$, we obtain singular values of $\mathbf{H}$ and $\mathbf{M}$ as $\left[\sigma^{\mathbf{H}}_1, \sigma^{\mathbf{H}}_2, \dots, \sigma^{\mathbf{H}}_d\right]$ and $\left[\sigma^{\mathbf{M}}_1, \sigma^{\mathbf{M}}_2, \dots, \sigma^{\mathbf{M}}_d\right]$. 

The fast singular value decay implies that the first singular value is significantly larger than others, \textit{i.e.,} $\sigma^{\mathbf{H}}_1\gg\sigma^{\mathbf{H}}_2$ and $\sigma^{\mathbf{M}}_1\gg\sigma^{\mathbf{M}}_2$. It further implies the insufficient diversity of sequence and item embeddings because the determinant~(re-scaled sum of singular values) of the embedding, \textit{i.e.,} the diversity from the Eq.~(\ref{eq:diverse_det}), achieves a small value when the fast singular value decay happens (most singular values are small values).

\section{Singular Spectrum Smoothing}
Motivated by the theoretical connection between the representation degeneration and diversity via the determinant of embedding, we present our proposed \modelname framework with a novel metric to measure the degree of representation degeneration and a novel singular spectrum smoothing regularization for alleviating the degeneration issue and controlling the diversity simultaneously. 

\subsection{Degree of Spectrum Smoothness}
% It remains unknown how to measure the degree of the degeneration issue in the embeddings. 
We propose a simple, intuitive, and straightforward metric to measure the degree of singular spectrum smoothness, which is calculated as the \textbf{A}rea \textbf{U}nder \textbf{S}ingular value \textbf{C}urve~(AUSC). 
The area under singular curve~(AUSC) is calculated as the sum of normalized singular values as follows:
% We formulate the AUSC as follows:
\begin{align}
\label{eq:degree_smooth}
    \text{AUSC} = \sum_{i=1}^k\frac{\sigma_i}{\sigma_{\text{max}}}.
\end{align}
When the singular value curve decays fast, as observed in the degenerated examples in Fig.~(\ref{fig:motivation_singular}), the area under the singular curve~(AUSC) is typically small because singular values other than the largest one are all small and all singular values are normalized. We can also see that when all singular values approximate the largest singular value, \textit{i.e.,} $|\sigma_i-\sigma_{\text{max}}|\approx 0$ for all $\sigma_i$, the AUSC is maximized, in which the singular curve becomes flatter. For example, our toy example shown in Fig.~(\ref{fig:intuition_singular}) compares two singular curves, in which the left one is the fast decaying curve and the right one is an improved and more flat singular curve. The left fast decaying curve has an AUSC value of 13.64, while the right smoothed curve has an AUSC value of 26.73.

\subsection{Dual Smoothing}
We build upon the AUSC to propose a singular spectrum smoothing regularization as a surrogate of optimizing AUSC. 
With the discussion in Section~\ref{sec:degen_discussion}, item and sequence degeneration issues are intertwined. Thus, we deploy the proposed singular spectrum smoothing regularization to both sequence output and item embeddings, which is referred to as dual smoothing. Specifically, the singular spectrum smoothing losses for both sequence $\mathcal{L}_{\text{seq}}$ and item $\mathcal{L}_{\text{item}}$ are conducted as follows:
\begin{align}
\label{eq:smooth_loss}
    \mathcal{L}_{\text{seq}} = -\frac{||\mathbf{H}||_{*}}{||\mathbf{H}||_{F}}, \ \ \ \ \ \mathcal{L}_{\text{item}} = -\frac{||\mathbf{M}||_{*}}{||\mathbf{M}||_{F}},
\end{align}
where $\mathbf{H}$ is the sequence output embedding matrix of all users, and 
The singular spectrum smoothing loss is the lower bound of our proposed AUSC smoothness metric as follows:
\begin{align}
    \frac{||\mathbf{H}||_{*}}{||\mathbf{H}||_{F}}\leq \frac{||\mathbf{H}||_{*}}{\sigma_{\text{max}}^{\mathbf{H}}},\ \ \ \ \frac{||\mathbf{M}||_{*}}{||\mathbf{M}||_{F}}\leq \frac{||\mathbf{M}||_{*}}{\sigma_{\text{max}}^{\mathbf{M}}}.
\end{align}
The loss minimization of either $\mathcal{L}_{\text{seq}}$ or $\mathcal{L}_{\text{item}}$ achieves the joint effects of
\begin{itemize}[leftmargin=*]
    \item maximizing the nuclear norm~(\textbf{encouraging the sum of singular values}),
    \item minimizing the Frobenius norm~(\textbf{suppressing the largest singular value}),
    \item and enlarges more tail singular values.
\end{itemize}
% With Theorem~\ref{theorem:1}, the nuclear norm of $\mathbf{H}$ or $\mathbf{E}$ provides a lower bound on the rank~\cite{horn2012matrix}. 
When $\mathcal{L}_{\text{seq}}$ and $\mathcal{L}_{\text{item}}$ are minimized, the nuclear norm is maximized, improving the determinant and diversity of $\mathbf{H}$ and $\mathbf{M}$. 
We adopt the Frobenius norm instead of the largest singular value for more stable training. 
The minimization of $\mathcal{L}_{\text{seq}}$ and $\mathcal{L}_{\text{item}}$ maximizes the AUSC of sequence embeddings and item embeddings, \textit{i.e.,} more smooth singular curves are introduced. Specifically, when the singular spectrum smoothing loss is minimized, the Frobenius norm is minimized so that the infinite norm of item embedding $\mathbf{M}$ is alleviated. The largest singular value is suppressed. 

\textbf{Computational Efficiency:} the computational complexity of singular values calculation can be reduced when $\mathcal{L}_{\text{seq}}$ and $\mathcal{L}_{\text{item}}$ are conducted in batch, which is $O(\text{min}(B^2d, Bd^2))$~\cite{DBLP:conf/cvpr/CuiWZLH020}, and $B$ denotes the batch size and $d$ is the dimension size. Different from~\cite{DBLP:journals/corr/abs-2107-06154, DBLP:conf/cvpr/CuiWZLH020, chen2022nuclear}, we present the nuclear norm divided by the Frobenius norm on embeddings of sequence and item sides while only nuclear norm of the prediction outputs is maximized in~\cite{DBLP:journals/corr/abs-2107-06154, DBLP:conf/cvpr/CuiWZLH020, chen2022nuclear}.

\subsection{Optimization with Dual Smoothing}
With singular spectrum smoothing on both sequence output embeddings side $\mathcal{L}_{\text{seq}}$ and item side $\mathcal{L}_{\text{item}}$, the final optimization loss $\mathcal{L}$ with the recommendation loss in Eq.~(\ref{eq:rec_loss}) to be minimized is defined as follows:
\begin{align}
    \mathcal{L} = \mathcal{L}_{rec} + \lambda\mathcal{L}_{\text{seq}} + \beta\mathcal{L}_{\text{item}},
\end{align}
where $\lambda$ is the hyper-parameter for controlling the sequence output embeddings side smoothing weight, and $\beta$ is the hyper-parameter for controlling the item side smoothing weight contribution to the final loss minimization. The final recommendation list is generated by calculating the scores defined in Eq.~(\ref{eq:score_rec}) on all items for each user. The recommendation scores on all items are sorted in descending order to produce the top-$N$ list.

\section{Experiments}
We conduct several empirical studies to demonstrate the usefulness of the \modelname and its components. We also analyze the improvements for a better understanding of the underlying mechanism. Specifically, we answer the following six research questions~(\textbf{RQ}s):
\begin{itemize}[leftmargin=*]
    \item \textbf{RQ1:} Does \modelname provide more satisfactory recommendations than state-of-the-art baselines?
    \item \textbf{RQ2:} How does our item side spectrum smoothing $\beta$ affect the diversity and recommendation performance?
    \item \textbf{RQ3:} Does \modelname outperform existing degeneration solutions?
    \item \textbf{RQ4:} How does each component of \modelname contribute to the performance?
    % \item \textbf{RQ5:} Does \modelname alleviate the representation degeneration issue?
    \item \textbf{RQ5:} Where do improvements of \modelname come from?% (in Appendix)
\end{itemize}

\begin{table}[]
\centering
\caption{Datasets Statistics.}
\label{tab:data_stat}
\resizebox{0.48\textwidth}{!}{%
\begin{tabular}{@{}c|rrrrc@{}}
\toprule
Dataset & \multicolumn{1}{c}{\#users} & \multicolumn{1}{c}{\#items} & \multicolumn{1}{c}{\#interactions} & \multicolumn{1}{c}{density} & \begin{tabular}[c]{@{}c@{}}avg. \\ interactions \\ per user\end{tabular} \\ \midrule
Home & 66,519 & 28,237 & 551,682 & 0.03\% & 8.3\\
Beauty & 22,363 & 12,101 & 198,502 & 0.05\% & 8.3 \\
Toys & 19,412 & 11,924 & 167,597 & 0.07\% & 8.6 \\
Tools & 16,638 & 10,217 & 134,476 & 0.08\% & 8.1 \\
Office & 4,905 & 2,420 & 53,258 & 0.44\% & 10.8 \\
Total & 129K & 74K & 1.10M & \multicolumn{1}{c}{-} & - \\ 
\bottomrule
\end{tabular}%
}
\end{table}
\subsection{Datasets}
We present the statistics of the dataset in Table \ref{tab:data_stat}. We evaluate all models in four benchmarks public Amazon review datasets\footnote {\url{http://deepyeti.ucsd.edu/jianmo/amazon/index.html}}. 
% In the Amazon reviews dataset, several categories of user-item interactions with timestamps are available. 
We choose \textit{Beauty}, \textit{Toys and Games}~(Toys), \textit{Tools and Home}~(Tools), and \textit{Office Products}~(Office) categories in our experiments as these four categories are widely adopted benchmark categories~\cite{tang2018personalized, kang2018self, sun2019bert4rec, fan2021modeling, li2020time, fan2022sequential}. 
We treat the existence of user-item reviews as positive user-item interactions. For each user, we sort the interacted items chronologically to form the user interaction sequence. The last interaction of the user sequence is used for testing and the second to last one for validation. We adopt the standard 5-core pre-processing setting on users~\cite{tang2018personalized, kang2018self, sun2019bert4rec, fan2021modeling, li2020time, fan2022sequential} to filter out users with less than five interactions. 
The detailed dataset statistics are presented in Table~\ref{tab:data_stat}.

\subsection{Baselines and Hyper-parameter Grid Search}
We compare the proposed \modelname with state-of-the-art sequential recommendation methods. We only include the static method BPRMF~\cite{rendle2012bpr} due to the page limitation. For sequential methods, we compare Caser~\cite{tang2018personalized}, SASRec~\cite{kang2018self}, DT4SR~\cite{fan2021modeling}, BERT4Rec~\cite{sun2019bert4rec}, FMLP-Rec~\cite{zhou2022filter}, and STOSA~\cite{fan2022sequential}. DuoRec~\cite{qiu2022contrastive} addresses the degeneration issue in contrastive learning for SR. Note that we use only \textbf{one training negative sample} (\textit{i.e., } $N^{-}_u=1$) for models with the Cross-Entropy loss for fair comparisons~\cite{ding2019reinforced, chen2017sampling, mao2021simplex}.

\subsection{All Items Ranking and Diversity Evaluation}
We generate the top-N recommendation list for each user based on the dot-product between sequence output embedding and the entire item set in descending order. We \textbf{rank all items} for all models so that no item sampling bias is introduced in evaluation for fair comparisons~\cite{krichene2020sampled}. The evaluation includes classical top-N ranking metrics, Recall@N and NDCG@N. 
% Recall@N measures the percentage of sequences being recommended true preferred item in the top-N list. NDCG@N weights the ranking position of the ground truth preferred item in the ranking list by giving exponentially smaller weights to lower-ranked items.
We present the averaged results over all test users. The test results are reported based on the best validation results. We report metrics in multiple Ns, including $N=\{5, 10, 40\}$, which are widely adopted in most existing SR methods~\cite{kang2018self, sun2019bert4rec, fan2022sequential}. 

We evaluate the diversity of the recommendation list for all test users in two widely adopted metrics, including the Intra-list diversity~\cite{zhang2008avoiding, ziegler2005improving} and the Coverage@100 of item categories for each user's top-N recommendation list. We focus on the top-10 ranking list for intra-list diversity calculation and top-100 for the coverage metric. The intra-list diversity is calculated as follows:
\begin{align}
    Diversity(u) = 1 - \frac{\sum_{i\in\mathcal{I}_{u}^{N}}\sum_{j\in\mathcal{I}_{u}^{N}}\text{cos}(\mathbf{M}_i, \mathbf{M}_j)}{|\mathcal{I}_{u}^{N}|*|\mathcal{I}_{u}^{N}|},
\end{align}
where $\mathcal{I}_{u}^{N}$ denotes the top-$N$ recommendation list for the user $u$, $\text{cos}(\cdot, \cdot)$ represents the cosine similarity, $\mathbf{M}$ is the item embedding table defined in Section~\ref{sec:transformer_basis}, and the reported diversity is the average diversity over all users. The larger value of diversity indicates higher intra-list diversity for the recommendation. 
Note that, we also found similar improvements in other larger $N$s, such as $N=\{20, 40\}$.

% Please add the following required packages to your document preamble:
% \usepackage{booktabs}
% \usepackage{multirow}
% \usepackage{graphicx}
% \usepackage[normalem]{ulem}
% \useunder{\uline}{\ul}{}
\begin{table*}[]
\centering
\caption{Overall Performance Comparison Table. The best results are bold and the best baseline results are underlined, respectively. `Imp. vs. SASRec' represents the relative improvement against SASRec without spectrum smoothing. `Improve.' indicates the relative improvement against the best baseline performance.}
\label{tab:overall_perf}
\resizebox{\textwidth}{!}{%
\begin{tabular}{@{}c|ccccccccc|ccc@{}}
\toprule
Dataset & Metric & BPRMF & Caser & SASRec & DT4SR & BERT4Rec & FMLP-Rec & STOSA & DuoRec & \modelname & Imp. vs. SASRec & Improv. \\ \midrule
\multirow{6}{*}{Home} & Recall@5  & 0.0096 & OOM   & 0.0127 & 0.0129 & 0.0105   & 0.0118   & {\ul 0.0133} & 0.0130 & \textbf{0.0171} & +34.65\%        & +28.68\% \\
& NDCG@5    & 0.0062 & OOM   & 0.0087 & 0.0082 & 0.0067   & 0.0086   & {\ul 0.0093} & 0.0088 & \textbf{0.0115} & +32.18\%        & +23.09\% \\
& Recall@10 & 0.0152 & OOM   & 0.0188 & 0.0193 & 0.0186   & 0.0177   & {\ul 0.0196} & 0.0185 & \textbf{0.0252} & +34.04\%        & +28.40\% \\
& NDCG@10   & 0.0080 & OOM   & 0.0107 & 0.0109 & 0.0093   & 0.0106   & {\ul 0.0113} & 0.0106 & \textbf{0.0140} & +30.84\%        & +23.89\% \\
& Recall@40 & 0.0355 & OOM   & 0.0432 & 0.0385 & {\ul 0.0470}   & 0.0430   & 0.0427 & 0.0424 & \textbf{0.0548} & +26.85\%        & +16.47\% \\
& NDCG@40   & 0.0126 & OOM   & 0.0161 & {\ul 0.0169} & 0.0156   & 0.0161   & 0.0165 & 0.0157 & \textbf{0.0207} & +28.57\%        & +22.03\% \\
& Diversity & 0.5914 & OOM   & 0.6213 & 0.6412 & 0.6351   & 0.6344   & {\ul 0.6771} & 0.6711 & \textbf{0.7034} & +13.21\%        & +3.88\%  \\
& Cov@100 & 72.644 & OOM & 74.799 & 76.041 & 75.663 & 79.518 & 78.881 & 75.776 & 76.676 & +2.51\% & - \\
\midrule
\multirow{6}{*}{Beauty} & Recall@5 & 0.0300 & 0.0309 & 0.0416 & 0.0449 & 0.0396 & 0.0469 & 0.0504 & {\ul 0.0505} & \textbf{0.0517} & +24.27\% & +2.38\% \\
 & NDCG@5 & 0.0189 & 0.0214 & 0.0274 & 0.0296 & 0.0257 & 0.0325 & \textbf{0.0351} & 0.0310 & {\ul 0.0345} & +25.91\% & -1.71\% \\
 & Recall@10 & 0.0471 & 0.0407 & 0.0633 & 0.0695 & 0.0595 & 0.0586 & {\ul 0.0707} & 0.0685 & \textbf{0.0745} & +17.69\% & +5.37\% \\
 & NDCG@10 & 0.0245 & 0.0246 & 0.0343 & 0.0375 & 0.0321 & 0.0362 & {\ul 0.0416} & 0.0375 & \textbf{0.0418} & +21.82\% & +0.18\% \\
 & Recall@40 & 0.1089 & 0.0863 & 0.1342 & {\ul 0.1384} & 0.1285 & 0.0977 & 0.1367 & 0.1364 & \textbf{0.1451} & +8.12\% & +4.84\% \\
 & NDCG@40 & 0.0383 & 0.0345 & 0.0503 & 0.0533 & 0.0476 & 0.0448 & {\ul 0.0565} & 0.0541 & \textbf{0.0577} & +14.71\% & +2.12\% \\ 
 & Diversity & 0.5875 & 0.5436 & 0.6302 & 0.6530 & 0.6509 & 0.6358 & 0.6631 & {\ul 0.6691} & \textbf{0.7154} & +13.51\% & +6.90\% \\
 & Cov@100 & 44.881 & 45.147 & 45.066 & 46.127 & 45.668 & 47.929 & 47.333 & 47.199 & 48.050 & +6.62\% & +0.03\% \\
 \midrule
 % & AUSC(User) & 14.7124 & 13.7814 & 19.5412 & 21.4771 & 18.7121 & 20.7541 & 23.7821 & {\ul 24.1084} & \textbf{26.3376} & +34.78\% & +9.25\% \\ 
\multirow{6}{*}{Tools} & Recall@5 & 0.0216 & 0.0129 & 0.0284 & 0.0289 & 0.0189 & 0.0195 & {\ul 0.0312} & 0.0304 & \textbf{0.0350} & +23.23\% & +12.18\% \\
 & NDCG@5 & 0.0139 & 0.0091 & 0.0194 & 0.0196 & 0.0123 & 0.0166 & {\ul 0.0217} & 0.0201 & \textbf{0.0238} & +22.68\% & +9.68\% \\
 & Recall@10 & 0.0334 & 0.0193 & 0.0427 & 0.0430 & 0.0319 & 0.0313 & {\ul 0.0468} & 0.0401 & \textbf{0.0513} & +20.14\% & +9.62\% \\
 & NDCG@10 & 0.0177 & 0.0112 & 0.0240 & 0.0246 & 0.0165 & 0.0204 & {\ul 0.0267} & 0.0234 & \textbf{0.0290} & +20.83\% & +8.61\% \\
 & Recall@40 & 0.0715 & 0.0680 & {\ul 0.0879} & 0.0861 & 0.0778 & 0.0785 & 0.0861 & 0.0867 & \textbf{0.1047} & +19.11\% & +19.11\% \\
 & NDCG@40 & 0.0262 & 0.0235 & 0.0343 & 0.0342 & 0.0268 & 0.0271 & {\ul 0.0356} & 0.0345 & \textbf{0.0410} & +19.53\% & +15.17\% \\ & Diversity & 0.5663 & 0.5223 & 0.6108 & 0.6353 & 0.6389 & 0.6429 & {\ul 0.6549} & 0.6438 & \textbf{0.7494} & +22.69\% & +14.43\% \\ 
  & Cov@100 & 60.781 & 63.473 & 62.871 & 67.134 & 66.743 & 68.545 & 67.441 & 67.851 & 68.212 & +8.50\% & - \\
  \midrule
 % & AUSC(User) & 15.7481 & 15.5416 & 17.0987 & 18.1241 & 17.0147 & 17.7161 & {\ul 18.9104} & 18.3715 & \textbf{20.7698} & +21.47\% & +9.83\% \\ 
\multirow{6}{*}{Toys} & Recall@5 & 0.0301 & 0.0240 & 0.0551 & 0.0550 & 0.0300 & {\ul 0.0625} & 0.0577 & 0.0580 & \textbf{0.0631} & +14.52\% & +0.96\% \\
 & NDCG@5 & 0.0194 & 0.0210 & 0.0377 & 0.0360 & 0.0206 & {\ul 0.0423} & 0.0412 & 0.0401 & \textbf{0.0431} & +14.32\% & +1.89\% \\
 & Recall@10 & 0.0460 & 0.0262 & 0.0797 & {\ul 0.0835} & 0.0466 & 0.0820 & 0.0800 & 0.0784 & \textbf{0.0852} & +6.90\% & +2.04\% \\
 & NDCG@10 & 0.0245 & 0.0231 & 0.0465 & 0.0437 & 0.0260 & {\ul 0.0485} & 0.0481 & 0.0461 & \textbf{0.0503} & +8.17\% & +3.71\% \\
 & Recall@40 & 0.1007 & 0.0909 & 0.1453 & {\ul 0.1478} & 0.0982 & 0.1406 & 0.1469 & 0.1452 & \textbf{0.1492} & +2.68\% & +0.95\% \\
 & NDCG@40 & 0.0368 & 0.0346 & 0.0604 & 0.0590 & 0.0376 & {\ul 0.0617} & 0.0611 & 0.0607 & \textbf{0.0647} & +7.12\% & +4.86\% \\ & Diversity & 0.6005 & 0.5924 & 0.6635 & 0.6706 & 0.6789 & 0.6895 & {\ul 0.6948} & 0.6936 & \textbf{0.7275} & +9.65\% & +4.71\% \\ 
 & Cov@100 & 42.981 & 44.217 & 43.444 & 45.662 & 44.891 & 49.507 & 46.189 & 46.771 & 47.572 & +9.50\% & - \\
 \midrule
 % & AUSC(User) & 15.7127 & 15.5674 & 17.0014 & 18.1047 & 16.7514 & {\ul 19.6712} & 19.3101 & 19.2134 & \textbf{20.2535} & +19.13\% & +2.67\% \\ 
\multirow{6}{*}{Office} & Recall@5 & 0.0214 & 0.0302 & 0.0656 & 0.0630 & 0.0485 & 0.0508 & {\ul 0.0677} & 0.0665 & \textbf{0.0714} & +8.84\% & +5.47\% \\
 & NDCG@5 & 0.0144 & 0.0186 & 0.0428 & 0.0421 & 0.0309 & 0.0343 & {\ul 0.0461} & 0.0456 & \textbf{0.0489} & +14.25\% & +6.07\% \\
 & Recall@10 & 0.0306 & 0.0550 & 0.0989 & 0.0940 & 0.0848 & 0.0977 & {\ul 0.1021} & 0.1005 & \textbf{0.1036} & +4.75\% & +1.47\% \\
 & NDCG@10 & 0.0173 & 0.0266 & 0.0534 & 0.0521 & 0.0426 & 0.0497 & {\ul 0.0572} & 0.0556 & \textbf{0.0593} & +11.05\% & +3.67\% \\
 & Recall@40 & 0.0718 & 0.1549 & 0.2251 & 0.2186 & 0.2230 & 0.1836 & {\ul 0.2346} & 0.2271 & \textbf{0.2367} & +5.15\% & +0.89\% \\
 & NDCG@40 & 0.0266 & 0.0487 & 0.0815 & 0.0797 & 0.0729 & 0.0685 & {\ul 0.0858} & 0.0817 & \textbf{0.0888} & +8.96\% & +3.50\% \\ & Diversity & 0.5920 & 0.5797 & 0.6167 & 0.6681 & 0.6551 & 0.6766 & {\ul 0.6889} & 0.6857 & \textbf{0.7350} & +19.18\% & +6.19\% \\
 & Cov@100 & 49.887 & 50.243 & 51.299 & 51.341 & 51.001 & 51.335 & 51.661 & 51.514 & 52.733 & +2.87\% & +2.08\% \\
 \bottomrule
 % & AUSC(User) & 10.7215 & 12.0174 & 13.6663 & 13.5145 & 12.0071 & 13.7114 & 17.7214 & {\ul 18.5244} & \textbf{19.2431} & +40.81\% & +3.88\% \\ 
\end{tabular}%
}
\end{table*}

\subsection{Overall Comparisons~(RQ1)}
We present the overall performance comparison with relevant state-of-the-art baselines in Table~\ref{tab:overall_perf}. From Table~\ref{tab:overall_perf}, we can easily conclude the superiority of \modelname in providing sequential recommendations in various datasets and achieving not only better performance but also more diversity. We have several observations to conclude as follows:
\begin{itemize}[leftmargin=*]
    \item \textbf{\modelname achieves the best performance and diversity, and outperforms SASRec with large improvement margins.} In most performance metrics and all datasets, \modelname significantly outperforms the second-best baseline with 0.52\% to 12.33\% in ranking performance and 4.71\% to 14.41\% in diversity. The improvements demonstrate the effectiveness of \modelname in the sequential recommendation, while the recommendation diversity is also noticeably improved. Furthermore, as \modelname is built on top of the original implementation of SASRec~\cite{kang2018self}, we observe 6.92\% to 22.53\% relative improvements over SASRec. We argue that the improvements originate from the singular spectrum smoothings on both sequence and item sides.
    \item \textbf{\modelname can achieve better balance between recommendation quality and diversity.} Comparing the diversity of \modelname and baselines, we can observe that \modelname consistently achieves better diversity results while having better performances, indicating the better capability of \modelname in balancing the recommendation quality and diversity. 
    \item \textbf{\modelname achieves performance larger improvements in sparser datasets.} Comparing \modelname and SASRec, we can observe at least 17\% improvements in both Beauty and Tools categories datasets, which have lower densities of 0.05\% and 0.07\%, respectively. The recommendation diversity is also significantly improved, especially for tools with a 14.43\%  improvement. However, in Toys and Office datasets with higher densities of 0.08\% and 0.44\%, the improvements are at most 14\%. This observation demonstrates that the proposed singular spectrum smoothing \modelname benefits sparser datasets, especially for datasets with a large portion of short sequences.
    \item \textbf{The consideration of Frobenius norm benefits the recommendation and diversity.} STOSA adopts the sum of two L2 distances on mean and covariance embeddings and DuoRec utilizes the uniformity loss based on the exponential L2 distance. We can observe that STOSA and DuoRec both outperform the dot-product-based SASRec framework in recommendation quality and diversity. 
    % Unlike the dot product approach, L2 distance considers the Frobenius norm of embeddings, even if it does not constrain the Frobenius norm, which again is the upper bound of the largest singular value of embeddings. 
    This observation further demonstrates the need to consider the Frobenius norm in embedding learning. Moreover, \modelname performs better than STOSA and DuoRec, demonstrating the necessity of considering the regularization from the singular spectrum perspective.
\end{itemize}

\begin{figure*}
\begin{subfigure}[t]{0.23\textwidth}
    % \
    \includegraphics[width=\textwidth]{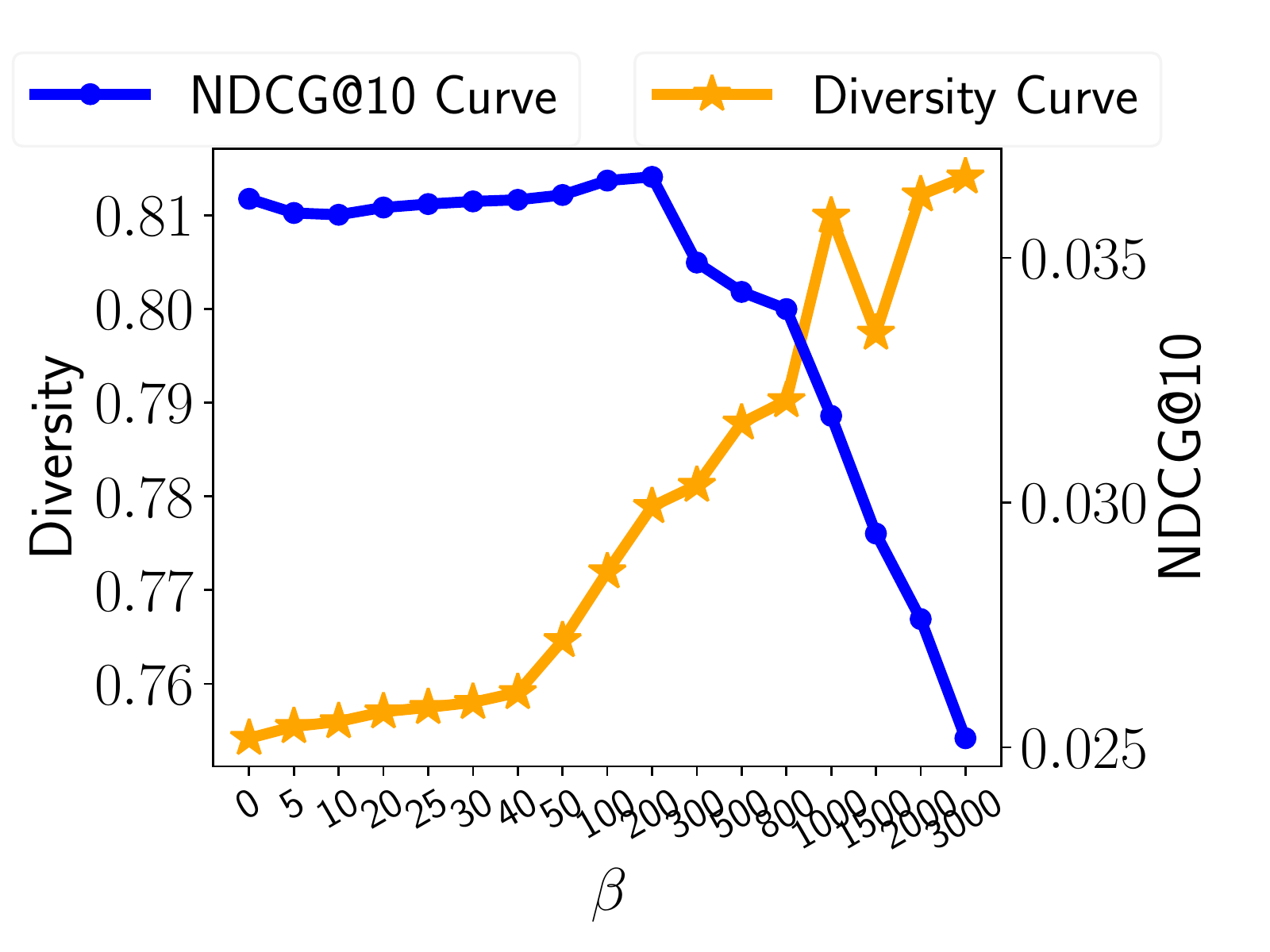}
    \caption{Beauty}
    \label{fig:beauty_item_diversity}
\end{subfigure}
\begin{subfigure}[t]{.23\textwidth}
    % \centering
    \includegraphics[width=\textwidth]{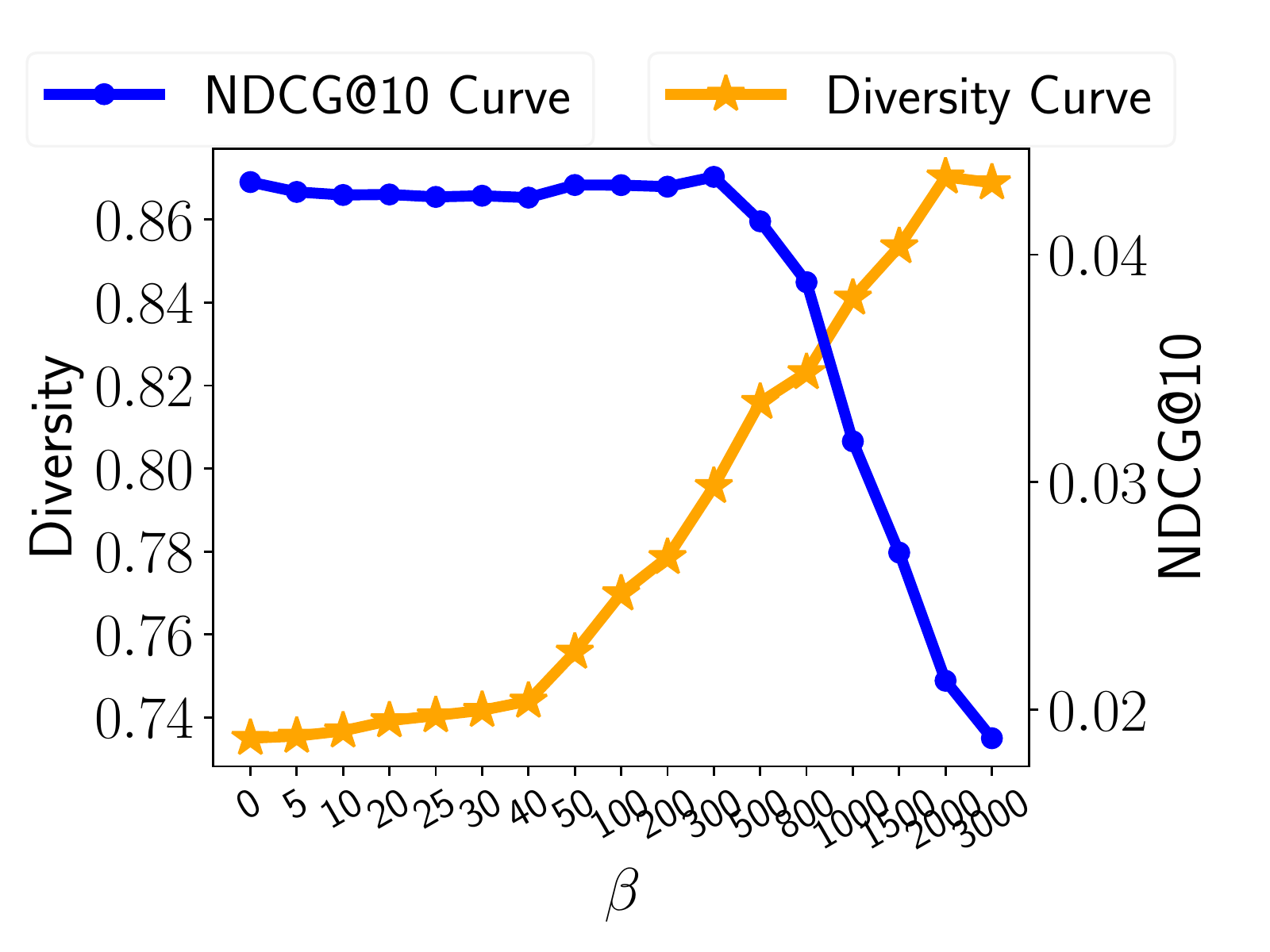}
    \caption{Toys}
    \label{fig:toys_item_diversity}
\end{subfigure}
% \vspace{-3mm}\\
% \\
\begin{subfigure}[t]{0.23\textwidth}
    % \
    \includegraphics[width=\textwidth]{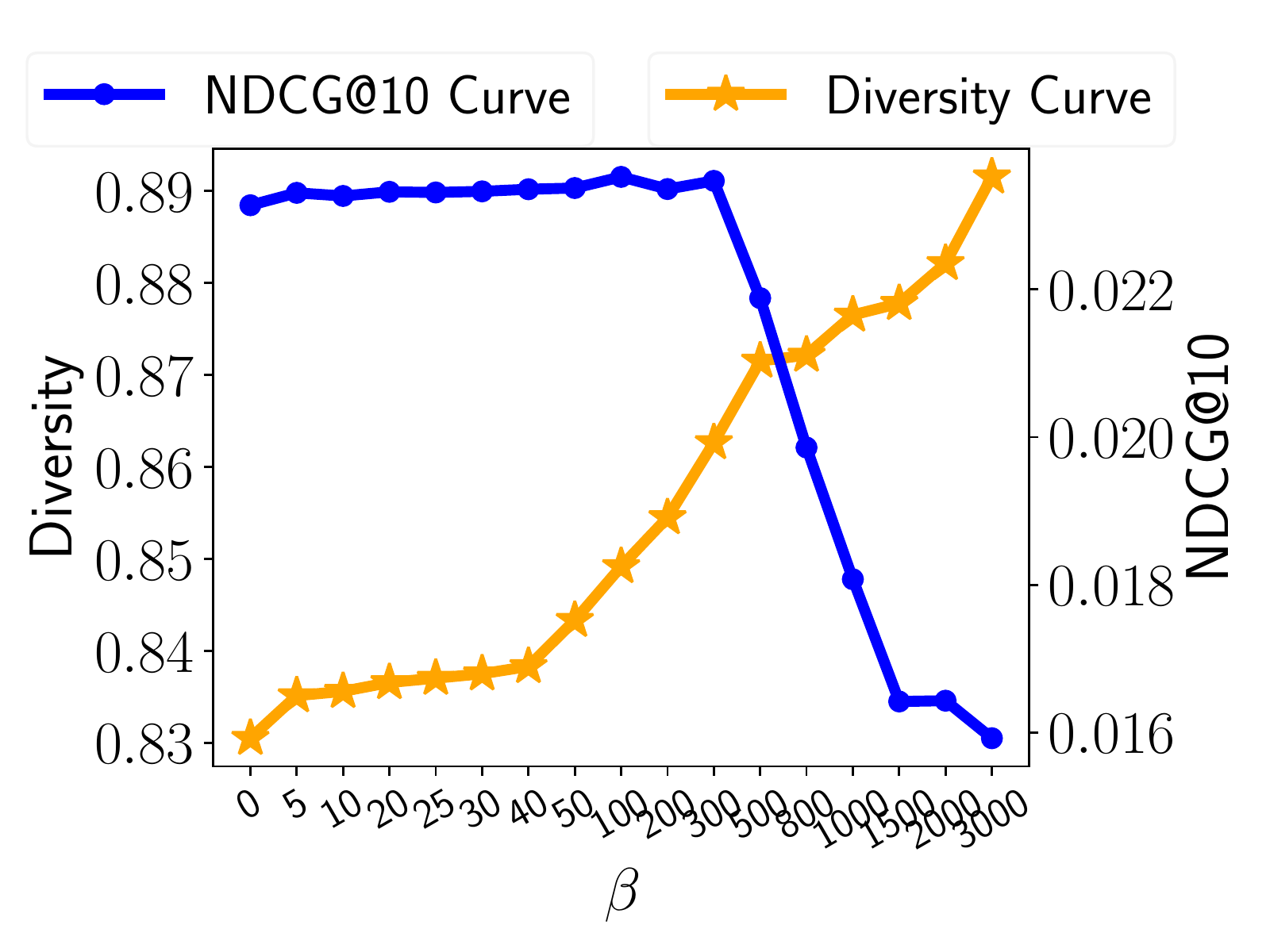}
    \caption{Tools}
    \label{fig:tools_item_diversity}
\end{subfigure}
\begin{subfigure}[t]{.23\textwidth}
    % \centering
    \includegraphics[width=\textwidth]{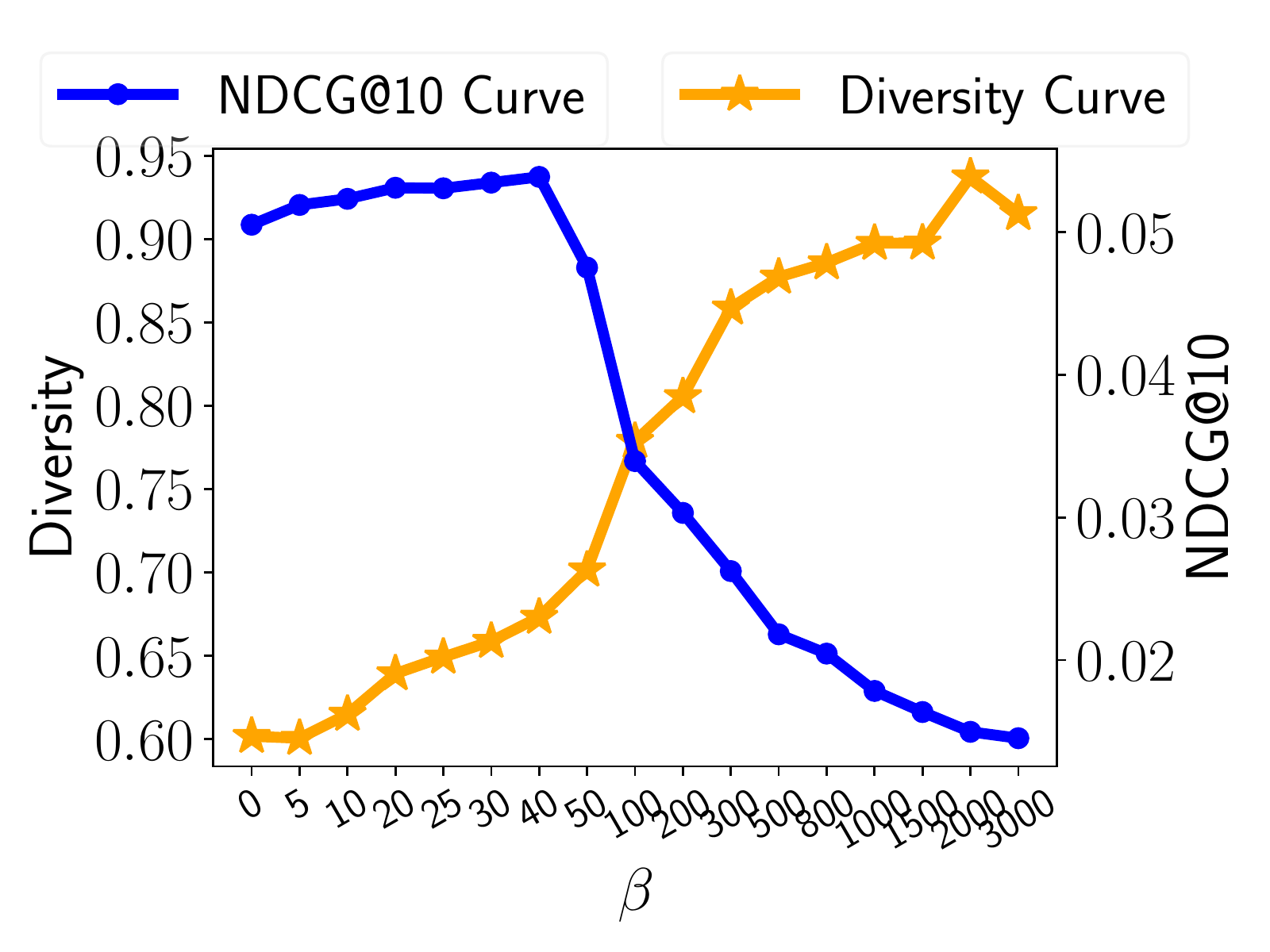}
    \caption{Office}
    \label{fig:office_item_diversity}
\end{subfigure}
% \vspace{-3mm}
\caption{Relationship between item spectrum smoothing weight $\beta$ and diversity based on embeddings.}
\label{fig:diversity}
\end{figure*}

\begin{figure*}
\begin{subfigure}[t]{0.23\textwidth}
    % \
    \includegraphics[width=\textwidth]{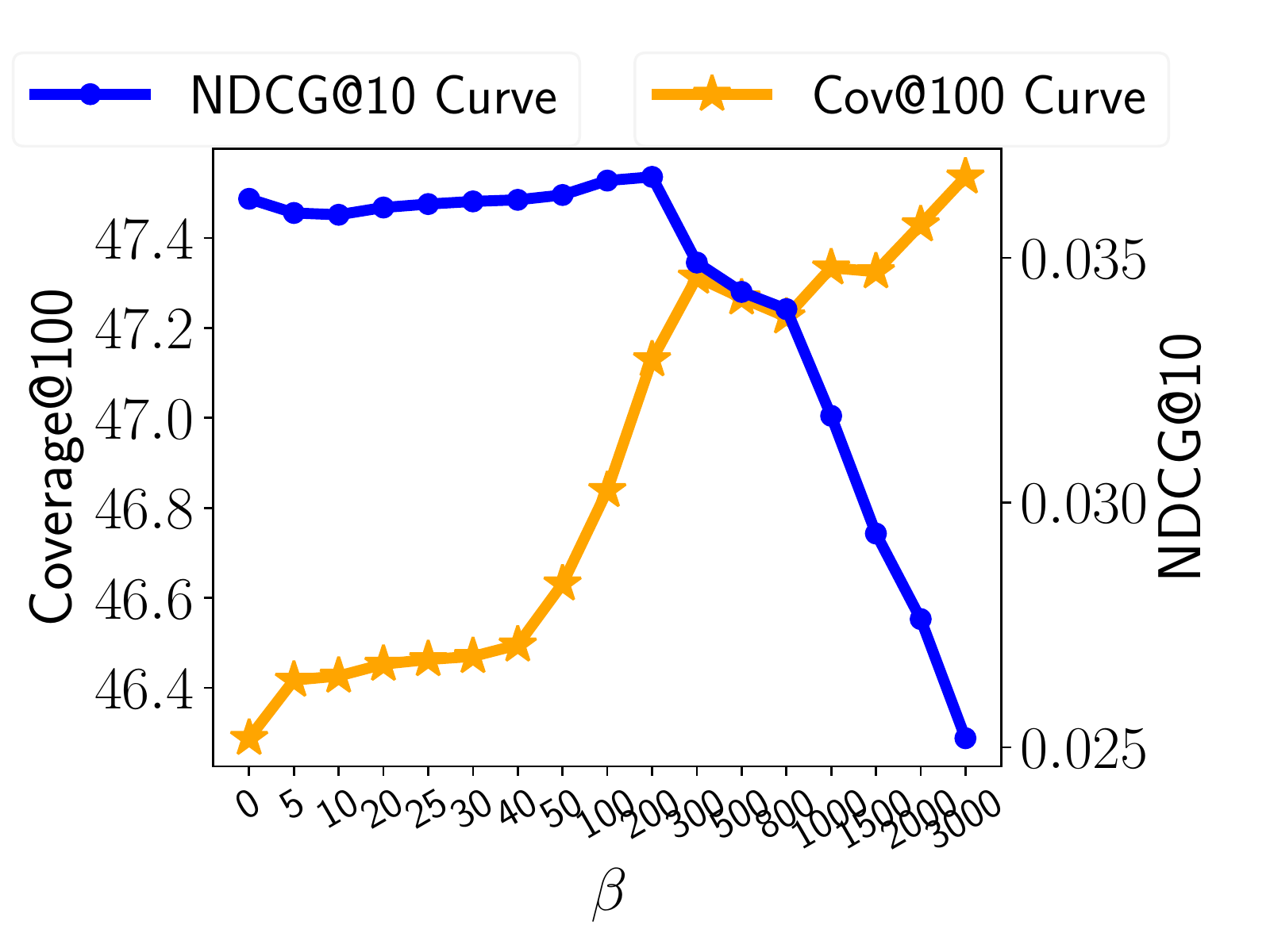}
    \caption{Beauty}
    \label{fig:beauty_cate_item_diversity}
\end{subfigure}
\begin{subfigure}[t]{.23\textwidth}
    % \centering
    \includegraphics[width=\textwidth]{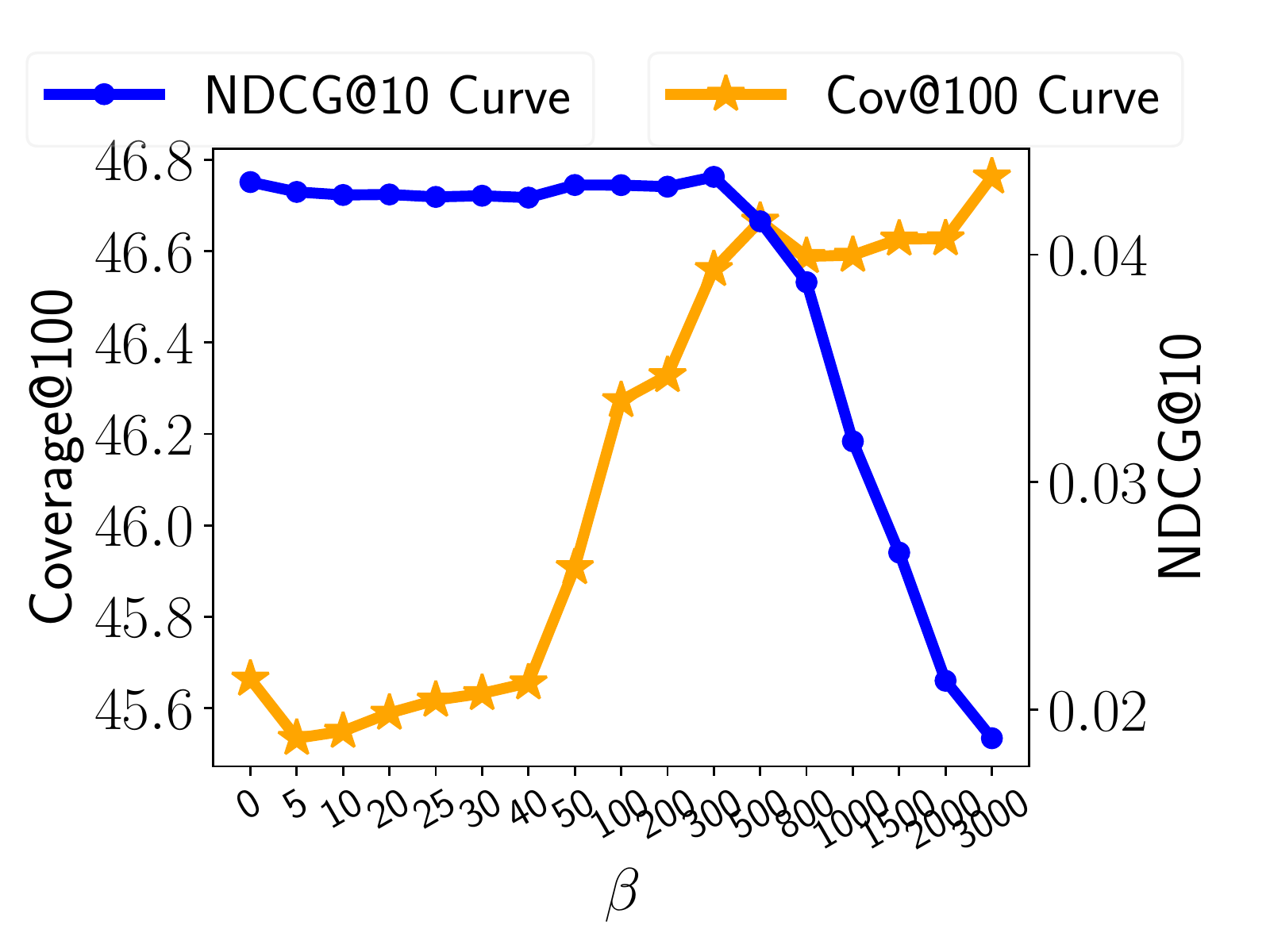}
    \caption{Toys}
    \label{fig:toys_cate_item_diversity}
\end{subfigure}
% \vspace{-3mm}\\
% \\
\begin{subfigure}[t]{0.23\textwidth}
    % \
    \includegraphics[width=\textwidth]{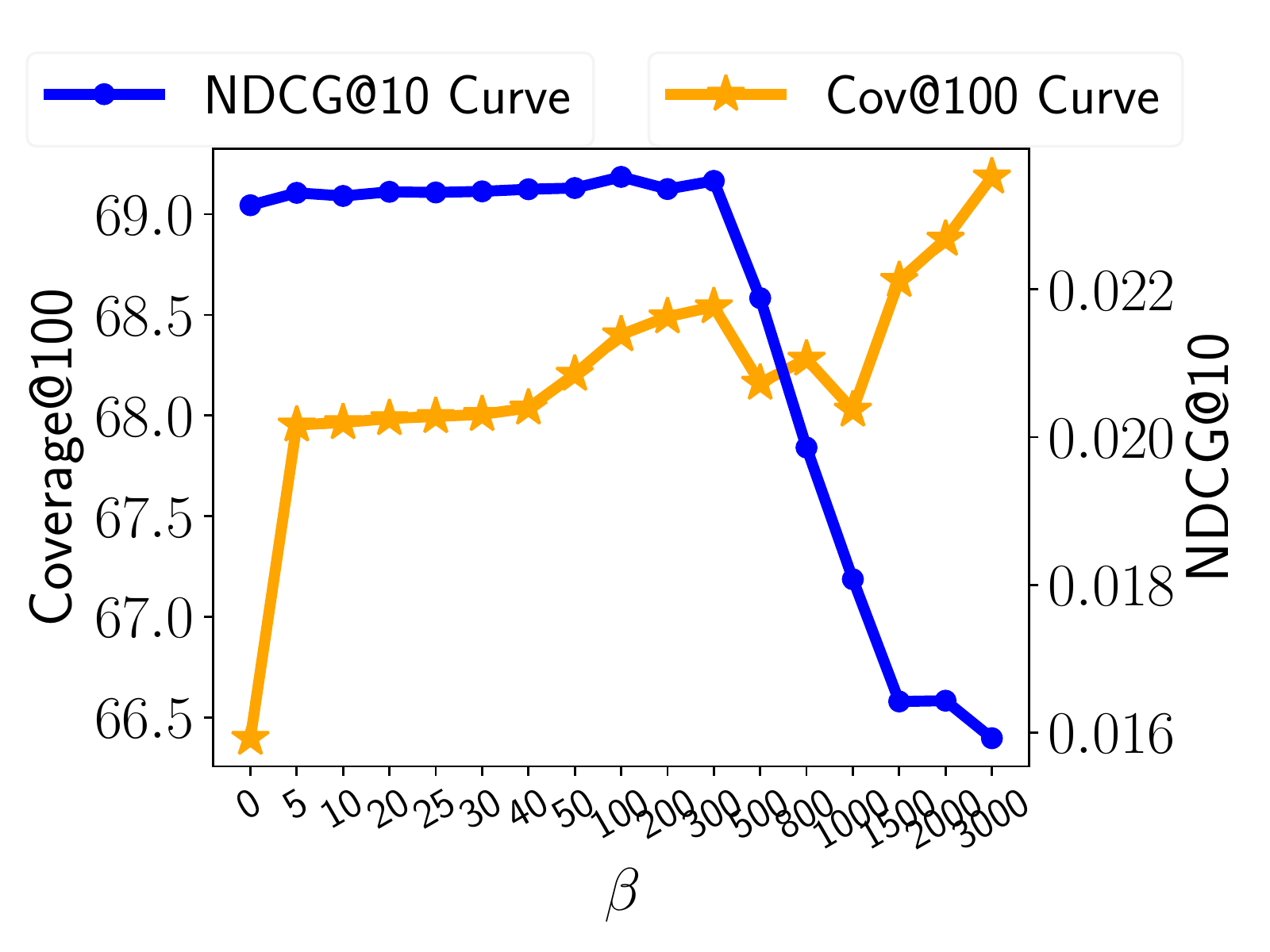}
    \caption{Tools}
    \label{fig:tools_cate_item_diversity}
\end{subfigure}
\begin{subfigure}[t]{.23\textwidth}
    % \centering
    \includegraphics[width=\textwidth]{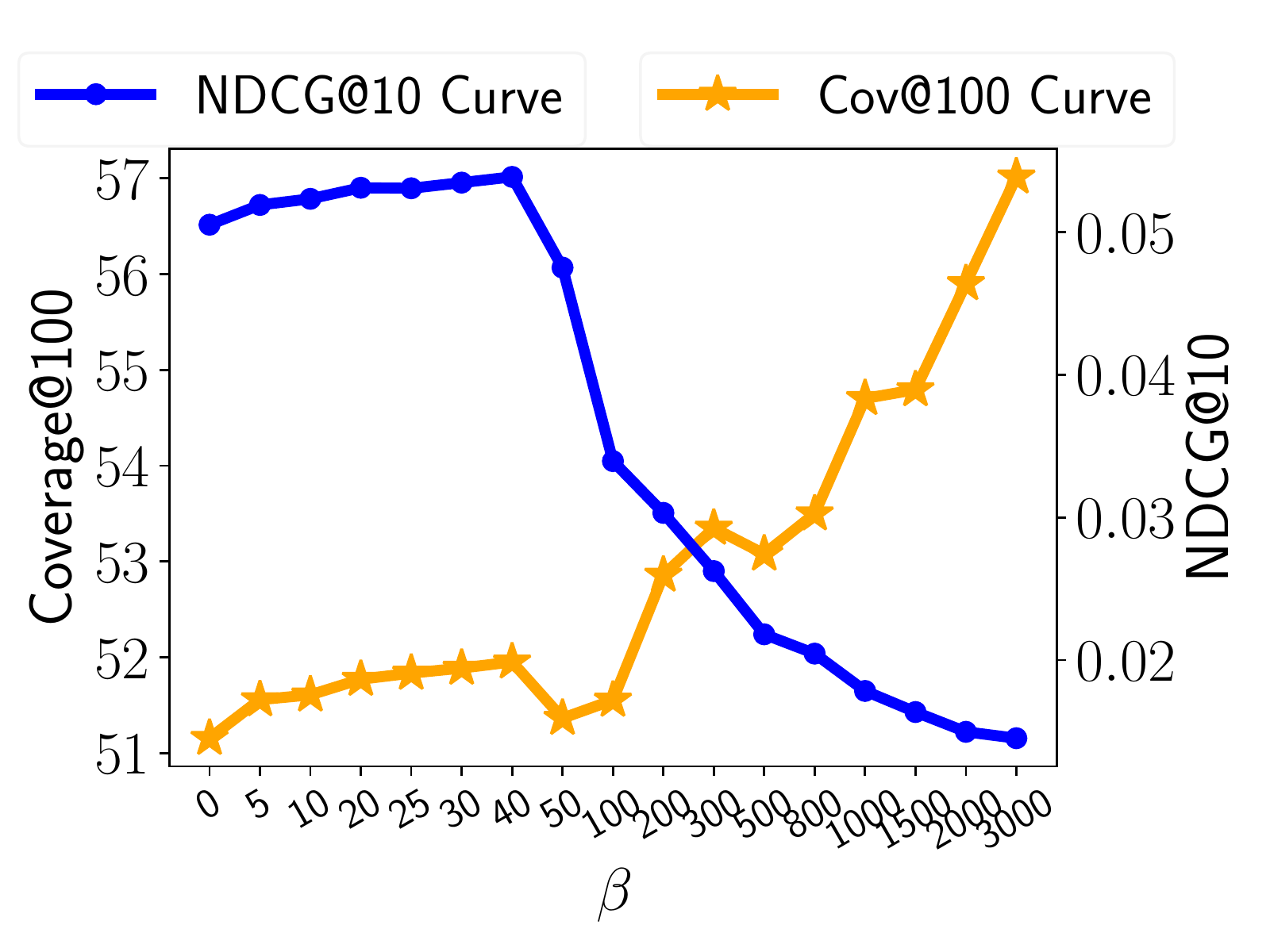}
    \caption{Office}
    \label{fig:office_cate_item_diversity}
\end{subfigure}
% \vspace{-3mm}
\caption{Relationship between item spectrum smoothing weight $\beta$ and diversity based on coverage@100.}
\label{fig:cate_diversity}
\end{figure*}

\subsection{Diversity and Item Smoothing $\mathcal{L}_{\text{item}}$~(RQ2)}
\label{sec:item_diversity}
\textbf{Item Spectrum Smoothing controlled by $\beta$ is closely related to the recommendation diversity.}
We demonstrate a strong connection between our item spectrum smoothing and the recommendation diversity by showing the relationship of $\beta$ and the diversity, as shown in Fig.~(\ref{fig:diversity}). In Fig.~(\ref{fig:diversity}), the x-axis denotes the value of $\beta$. As the $\beta$ becomes larger, we can observe that the diversity consistently increases, in all four datasets, demonstrating the strong connection between the recommendation diversity and our proposed item spectrum smoothing component. 

\textbf{Moreover, the recommendation performance drops when we demand too high recommendation diversity.} Another observation is that the NDCG@10 drops significantly when we set $\beta$ to a too-large value, which also indicates a higher diversity. This observation demonstrates that a proper selection of $\beta$ can achieve satisfactory recommendation performance and diversity simultaneously. However, demanding higher diversity can hurt the recommendation performance.

\begin{table}[]
\centering
\caption{Performance Comparison Table for various methods to alleviate degeneration issues.}
\label{tab:degen_perf}
\resizebox{0.48\textwidth}{!}{%
\begin{tabular}{@{}c|cccc|c@{}}
\toprule
Dataset & Metric & SASRec & SASRec+CosReg & SASRec+Euclidean & \multicolumn{1}{l}{\modelname} \\ \midrule
\multirow{4}{*}{Beauty} & Recall@5 & 0.0416 & 0.0410 & 0.0326 & 0.0517 \\
 & NDCG@5 & 0.0274 & 0.0250 & 0.0212 & 0.0345 \\
 & Recall@10 & 0.0633 & 0.0609 & 0.0487 & 0.0745 \\
 & NDCG@10 & 0.0343 & 0.0314 & 0.0264 & 0.0418 \\ \midrule
\multirow{4}{*}{Tools} & Recall@5 & 0.0284 & 0.0290 & 0.0230 & 0.0350 \\
 & NDCG@5 & 0.0194 & 0.0194 & 0.0145 & 0.0238 \\
 & Recall@10 & 0.0427 & 0.0435 & 0.0359 & 0.0513 \\
 & NDCG@10 & 0.0240 & 0.0240 & 0.0186 & 0.0290 \\ \midrule
\multirow{4}{*}{Toys} & Recall@5 & 0.0551 & 0.0457 & 0.0537 & 0.0631 \\
 & NDCG@5 & 0.0377 & 0.0302 & 0.0375 & 0.0431 \\
 & Recall@10 & 0.0797 & 0.0658 & 0.0717 & 0.0852 \\
 & NDCG@10 & 0.0465 & 0.0367 & 0.0432 & 0.0503 \\ \midrule
\multirow{4}{*}{Office} & Recall@5 & 0.0656 & 0.0612 & 0.0561 & 0.0714 \\
 & NDCG@5 & 0.0428 & 0.0400 & 0.0376 & 0.0489 \\
 & Recall@10 & 0.0989 & 0.0919 & 0.0862 & 0.1036 \\
 & NDCG@10 & 0.0534 & 0.0500 & 0.0472 & 0.0593 \\ \bottomrule
\end{tabular}%
}
\end{table}
\subsection{Degeneration Methods Comparison~(RQ3)}
\label{sec:degeneration_comparisons}
We compare the proposed \modelname with spectrum smoothing regularization with existing methods for alleviating the degeneration issue in word embeddings learning, including the cosine regularization~\cite{DBLP:conf/iclr/GaoHTQWL19} as follows:
\begin{align}
    \mathcal{L}_{\text{item}} = \frac{1}{|\mathcal{V}|\times|\mathcal{V}|}\sum_{i=1}^{|\mathcal{V}|}\sum_{j=1}^{|\mathcal{V}|}\frac{\mathbf{M}_{i}^{\top}\mathbf{M}_j}{||\mathbf{M}_{i}||\times||\mathbf{M}_{j}||},
\end{align}
and the Euclidean method~\cite{DBLP:conf/emnlp/0004GXMYS20} as follows:
\begin{align}
    \mathcal{L}_{\text{item}} &= -\frac{1}{|\mathcal{V}|\times|\mathcal{V}|}\sum_{i=1}^{|\mathcal{V}|}\sum_{j=1}^{|\mathcal{V}|}||\mathbf{M}_{i}-\mathbf{M}_j||_2. 
    % \nonumber \\ 
    % &=-\frac{1}{|\mathcal{V}|\times|\mathcal{V}|}\sum_{i=1}^{|\mathcal{V}|}\sum_{j=1}^{|\mathcal{V}|}||\mathbf{E}_{i}||_2-2\mathbf{E}_{i}^{\top}\mathbf{E}_{j}+||\mathbf{E}_j||_2.
\end{align}
We apply the cosine and Euclidean regularization methods to both sides. We report the best grid searched results of all regularization methods in Table~\ref{tab:degen_perf}. Note that the uniformity property adopted by DuoRec and DirectAU~\cite{qiu2022contrastive, DBLP:conf/kdd/WangYM000M22} falls into the Euclidean methods category.

As shown in Table~\ref{tab:degen_perf}, neither cosine regularization or Euclidean regularization can consistently outperform the original SASRec without any regularization. Arguably, the potential reason is that neither of these two methods constrains the Frobenius norm of embeddings, which is the upper bound of the largest singular value, as discussed in Section~\ref{sec:degen_discussion}. 
When cosine regularization and the Euclidean method are minimized, the naive solution is $||\mathbf{E}_i||_2\rightarrow \infty$ for all $i$ if we do not consider the supervised recommendation loss. Moreover, different from the word embedding task, sequences in the recommendation task are shorter on average, posing a greater challenge in addressing the degeneration in SR. However, our proposed singular spectrum smoothing method \modelname minimizes the Frobenius norm while maximizing the nuclear norm, achieving the goals of smoothing the singular value curve and suppressing the largest singular value simultaneously.

\begin{figure*}
\begin{subfigure}[t]{0.23\textwidth}
    % \
    \includegraphics[width=\textwidth]{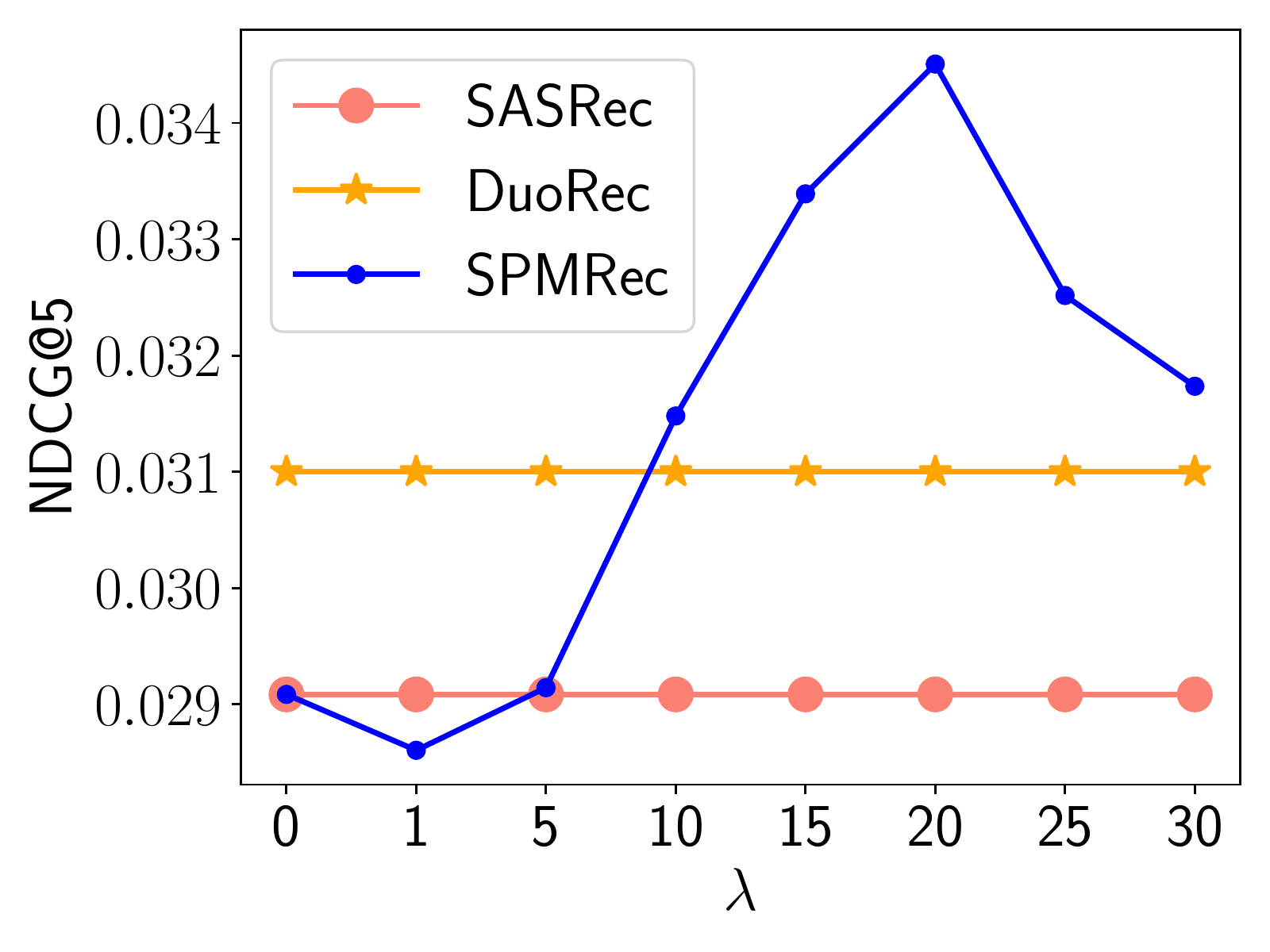}
    \caption{Beauty}
    \label{fig:beauty_item}
\end{subfigure}
\begin{subfigure}[t]{.23\textwidth}
    % \centering
    \includegraphics[width=\textwidth]{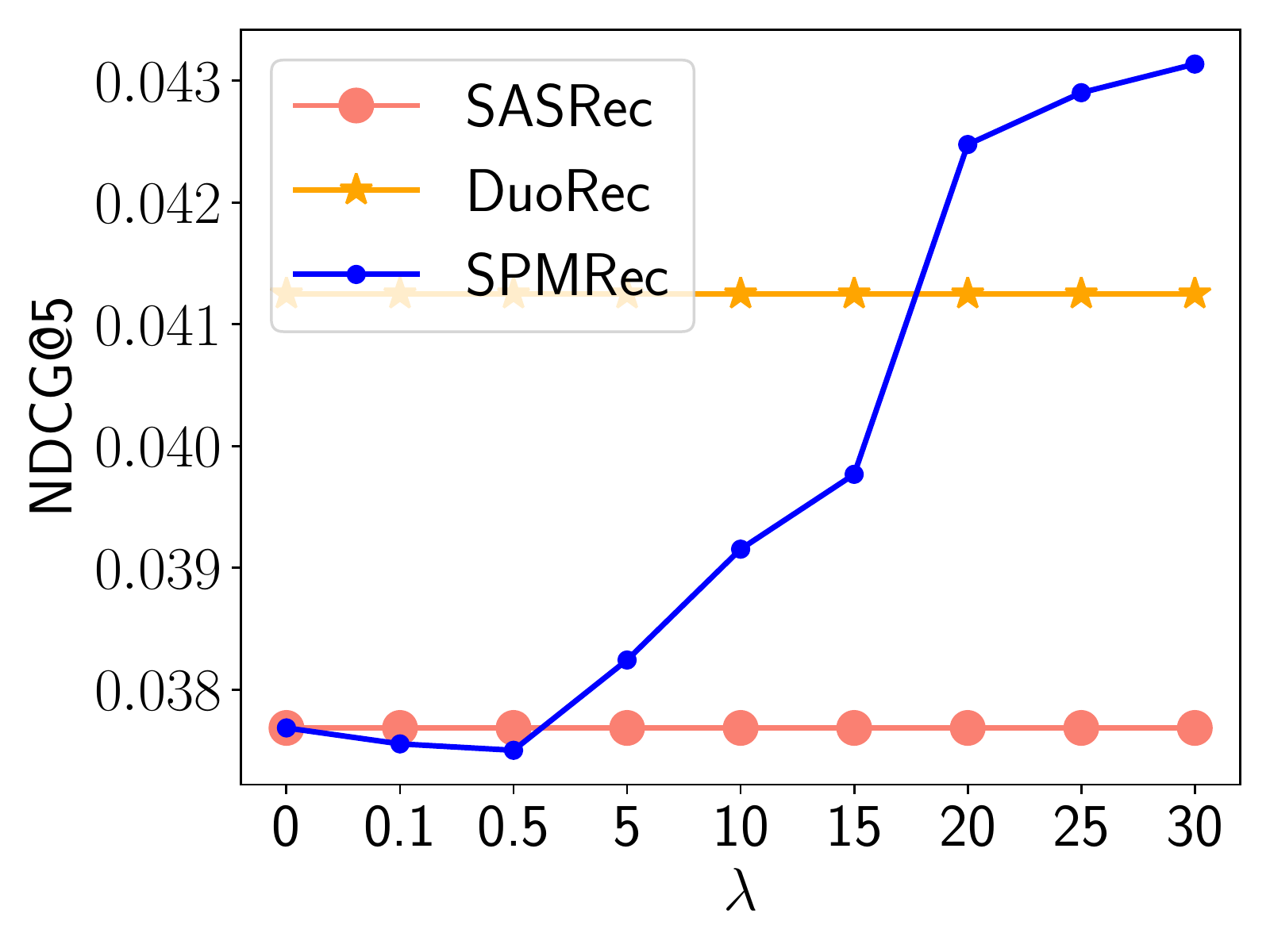}
    \caption{Toys}
    \label{fig:toys_item}
\end{subfigure}
% \vspace{-3mm}\\
% \\
\begin{subfigure}[t]{0.23\textwidth}
    % \
    \includegraphics[width=\textwidth]{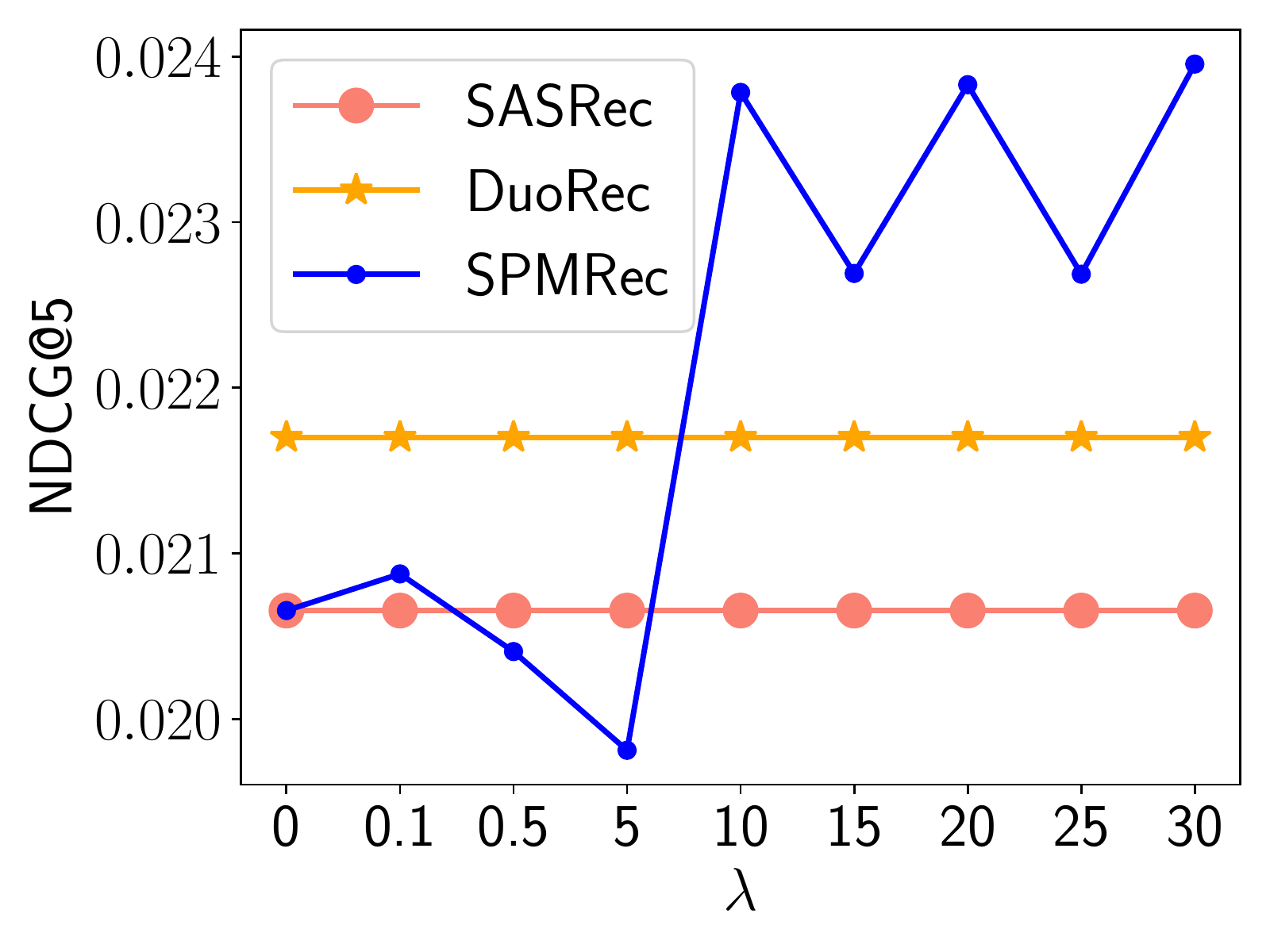}
    \caption{Tools}
    \label{fig:tools_item}
\end{subfigure}
\begin{subfigure}[t]{.23\textwidth}
    % \centering
    \includegraphics[width=\textwidth]{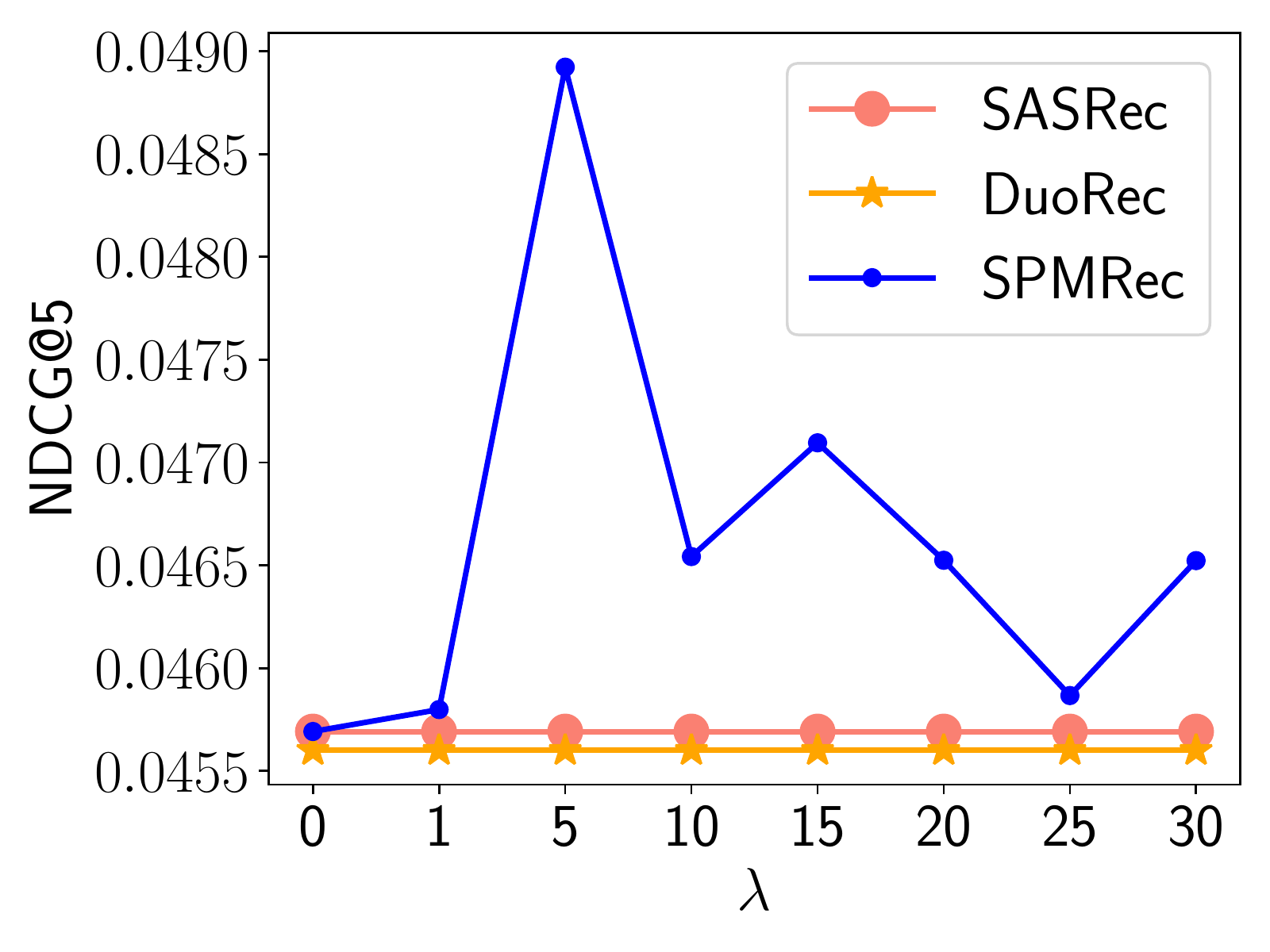}
    \caption{Office}
    \label{fig:office_item}
\end{subfigure}
% \vspace{-3mm}
\caption{NDCG@5 on different sequential spectrum smoothing weight $\lambda$ values. Best Viewed in colors.}
\label{fig:seq_smooth_lambda}
\end{figure*}

\begin{figure*}
\begin{subfigure}[t]{0.23\textwidth}
    % \
    \includegraphics[width=\textwidth]{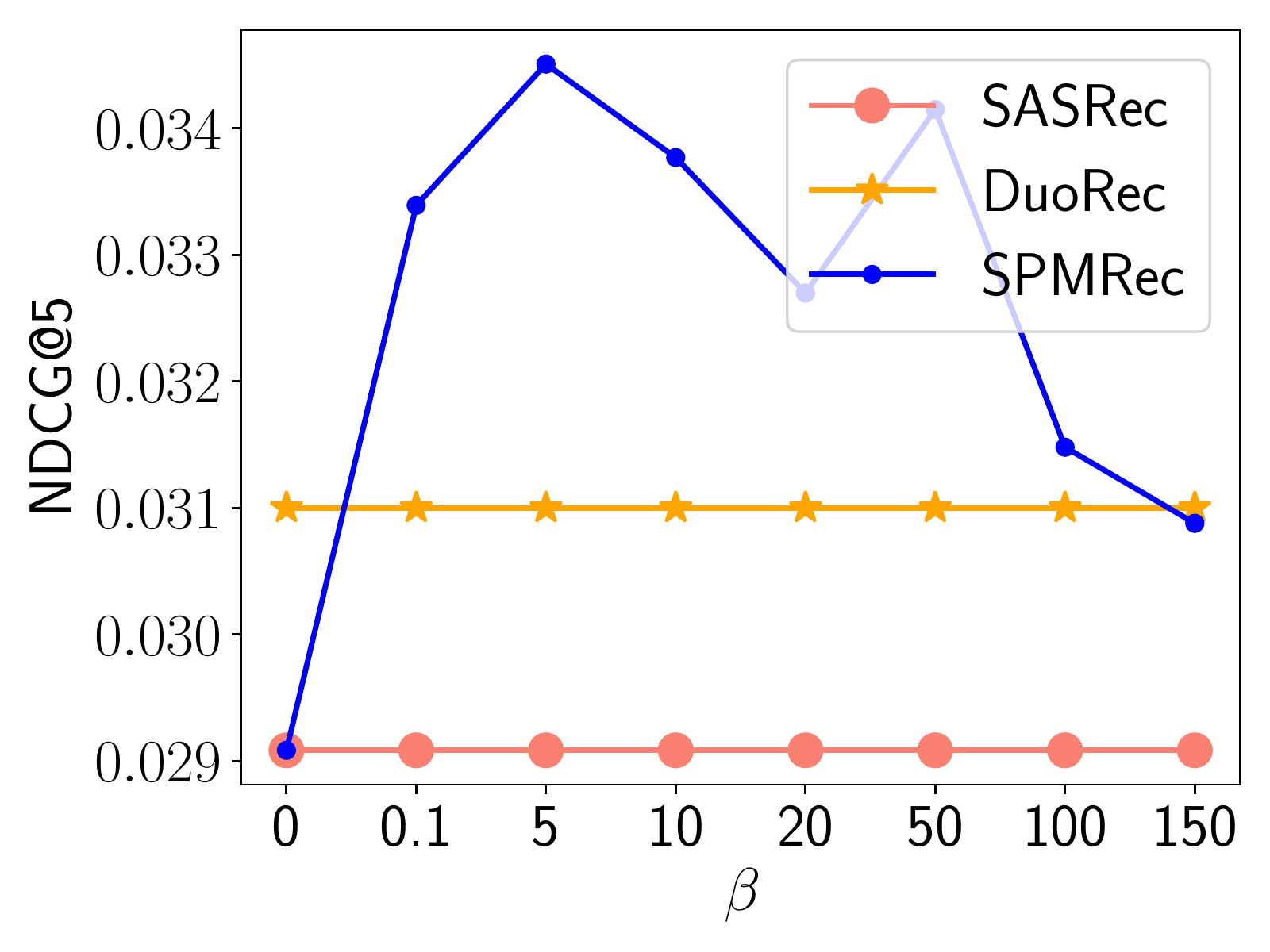}
    \caption{Beauty}
    \label{fig:beauty_item}
\end{subfigure}
\begin{subfigure}[t]{.23\textwidth}
    % \centering
    \includegraphics[width=\textwidth]{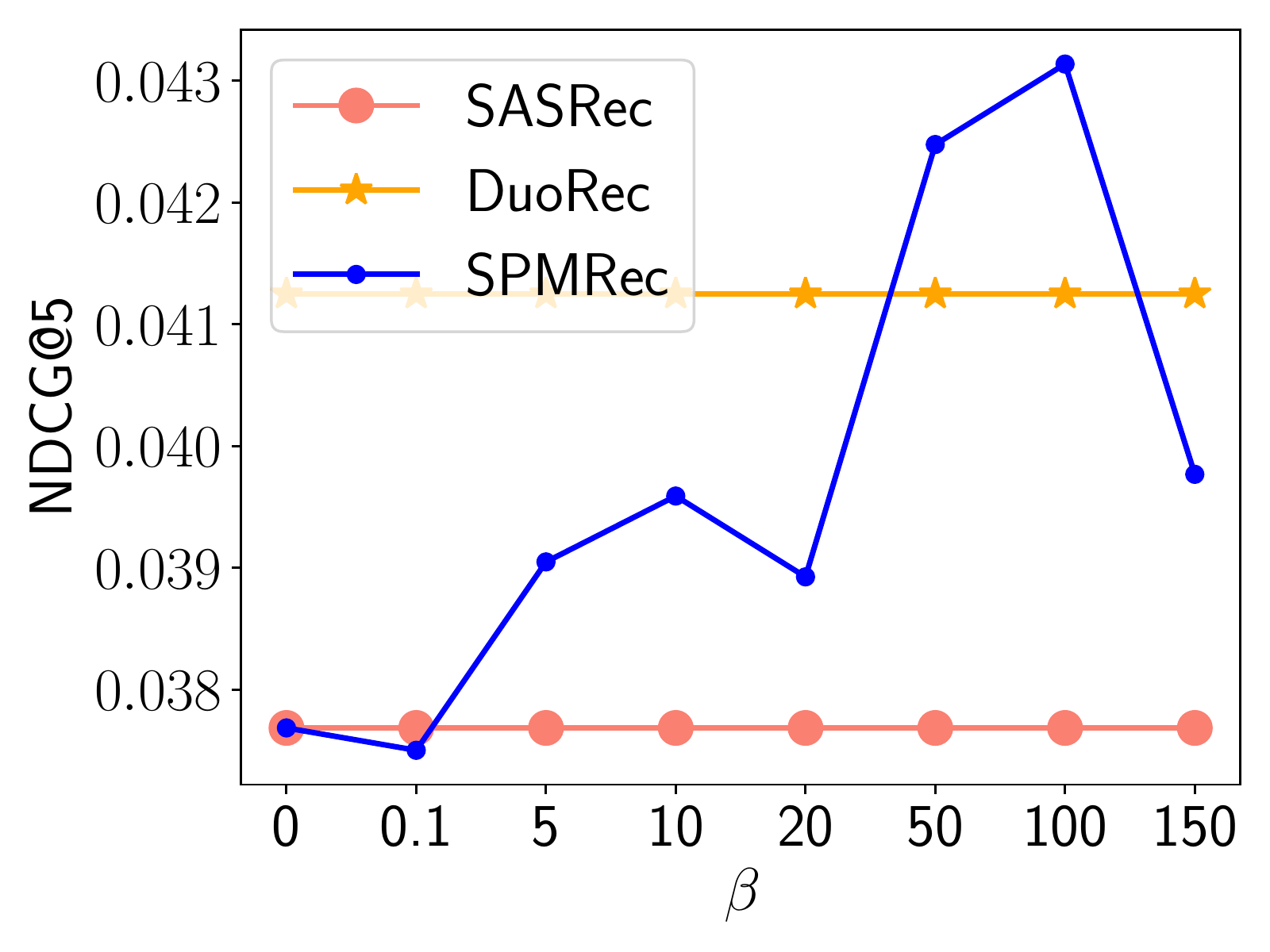}
    \caption{Toys}
    \label{fig:toys_item}
\end{subfigure}
% \vspace{-3mm}\\
% \\
\begin{subfigure}[t]{0.23\textwidth}
    % \
    \includegraphics[width=\textwidth]{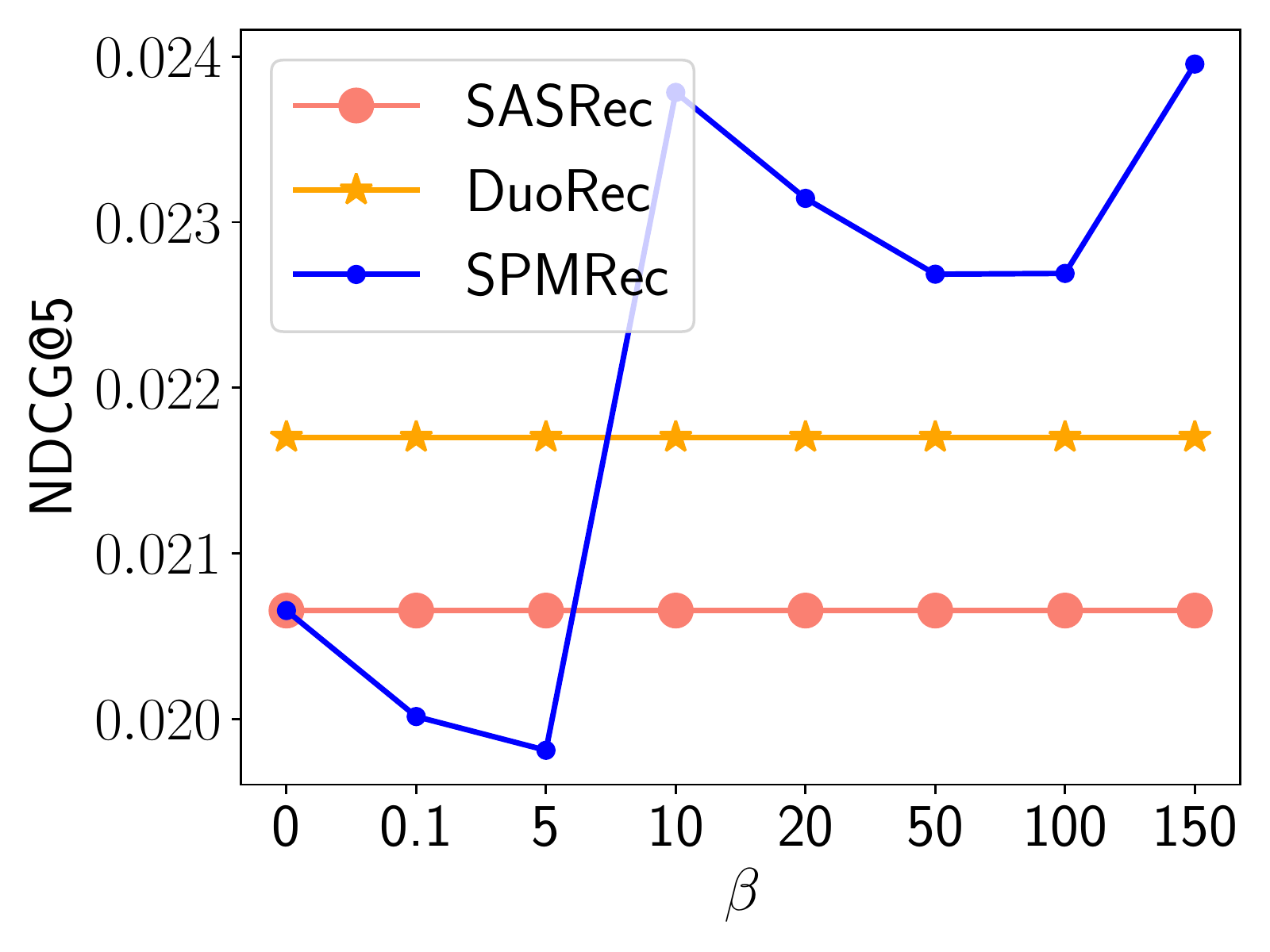}
    \caption{Tools}
    \label{fig:tools_item}
\end{subfigure}
\begin{subfigure}[t]{.23\textwidth}
    % \centering
    \includegraphics[width=\textwidth]{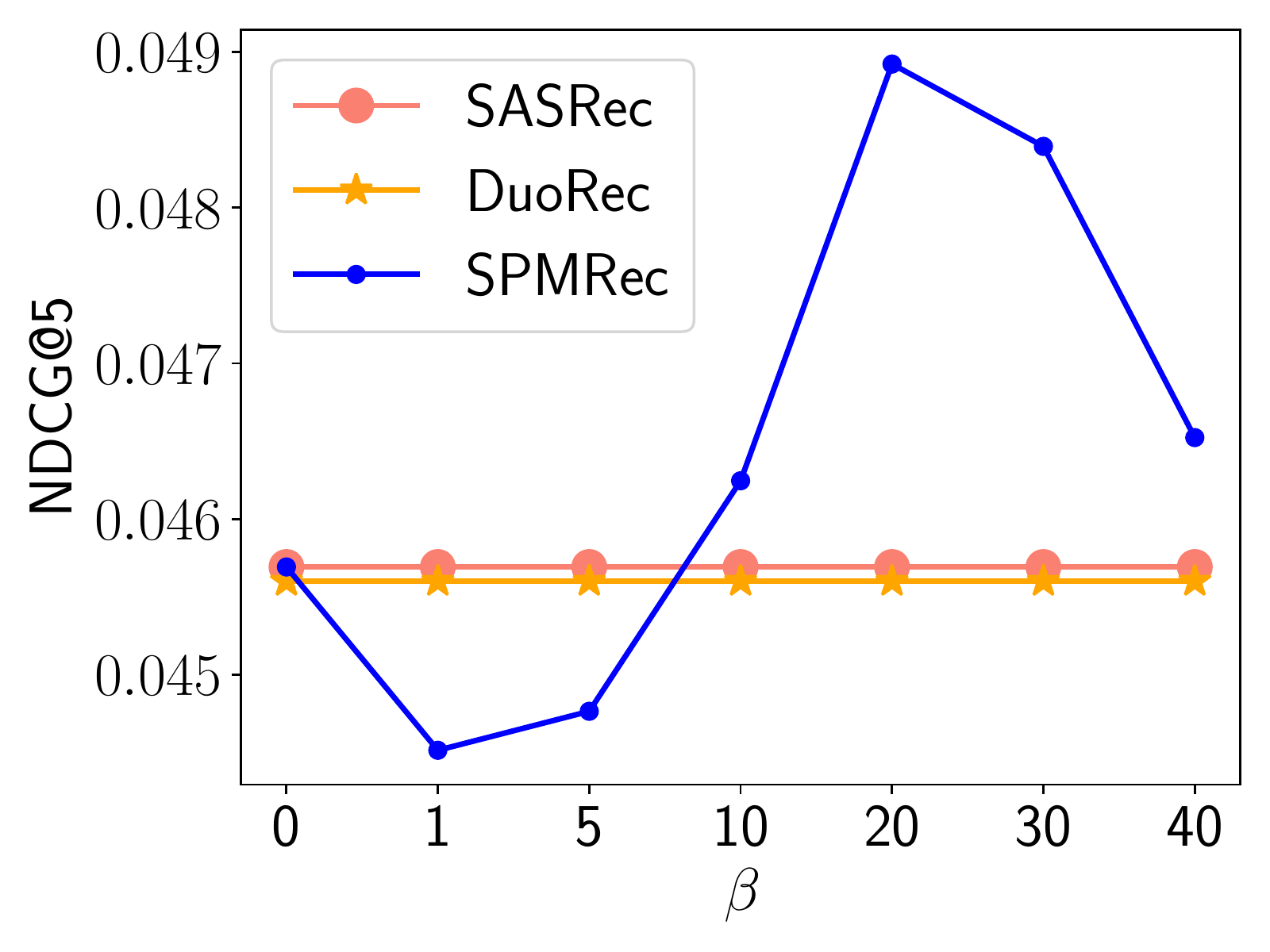}
    \caption{Office}
    \label{fig:office_item}
\end{subfigure}
% \vspace{-3mm}
\caption{NDCG@5 on different item spectrum smoothing weight $\beta$ values.}
\label{fig:item_smooth_beta}
\end{figure*}
\subsection{Ablation Study~(RQ4)}
We conduct the ablation study to show the effectiveness of two important components, including singular spectrum smoothing for sequence output and item embeddings. We investigate the smoothing weights $\lambda$~(for sequence output embeddings smoothing) and $\beta$~(for item embeddings smoothing) in Fig.~(\ref{fig:seq_smooth_lambda}) and Fig.~(\ref{fig:item_smooth_beta}), respectively. We include the SASRec~(without any regularization) and DuoRec~(with uniformity regularization). Note that the special case ($\lambda=0$ and $\beta=0$) is included, which is SASRec without any regularization. Additionally, the sequence outputs and item embeddings spectrum smoothing losses are in different scales ($10^{-1}$ for sequence embeddings and $10^{-4}$ for item embeddings). Hence, the scales of $\lambda$ and $\beta$ are different with the purpose of balancing the contributions of two regularizations. 

\subsubsection{Sequence Output Embedding Smoothing Weight $\lambda$}
We present the performance values with different values of $\lambda$ in Fig.~(\ref{fig:seq_smooth_lambda}) for all four datasets. Fig.~(\ref{fig:seq_smooth_lambda}) shows that the proper selection of $\lambda$ improves the performance significantly. Moreover, most $\lambda$ values achieve comparative and better performances than SASRec. These observations show that the sequence output embedding spectrum smoothing is beneficial for the sequential recommendation. 

\subsubsection{Item Embedding Smoothing Weight $\beta$}
The NDCG@5 values of different $\beta$ are shown in Fig.~(\ref{fig:item_smooth_beta}), which is controlling the spectrum smoothing component's contribution. The analysis is conducted in all four datasets used. Overall, when $\beta$ becomes larger, \textit{i.e.,} having a larger contribution to the overall loss, the performance increases. It shows the effectiveness of item embeddings singular spectrum smoothing for performance improvements.

\subsection{Improvements Analysis~(RQ5)}

\subsubsection{Improvements w.r.t. Sequence Length}
\begin{figure*}
\begin{subfigure}[t]{0.23\textwidth}
    % \
    \includegraphics[width=\textwidth]{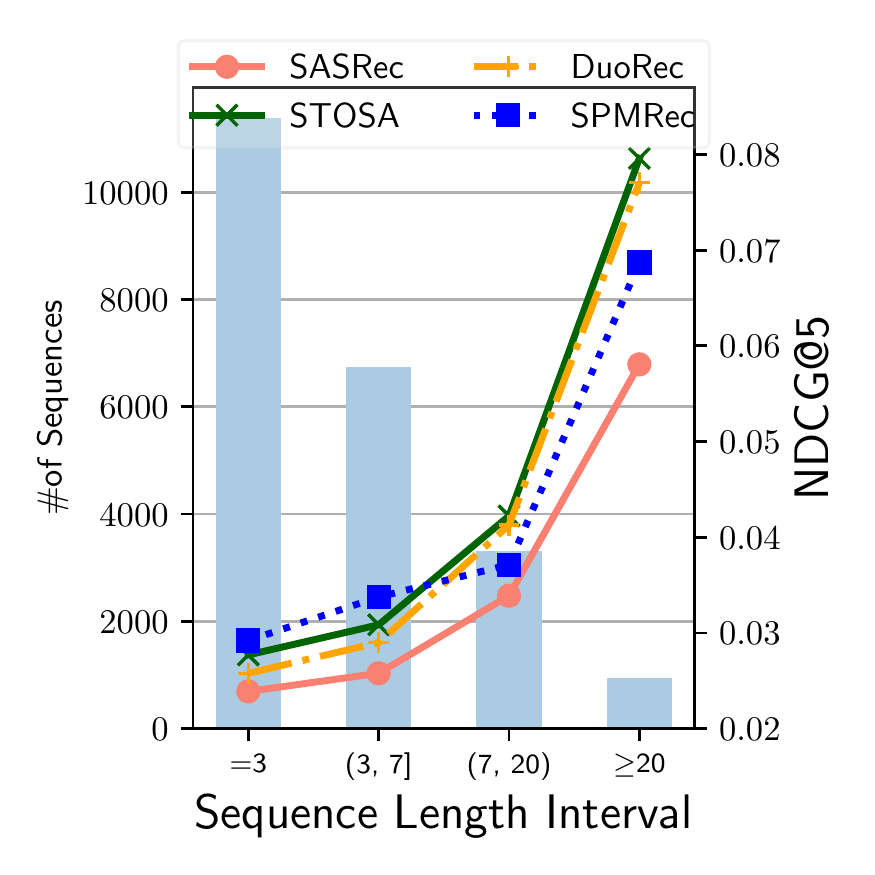}
    \caption{Beauty}
    \label{fig:beauty_item}
\end{subfigure}
\begin{subfigure}[t]{.23\textwidth}
    % \centering
    \includegraphics[width=\textwidth]{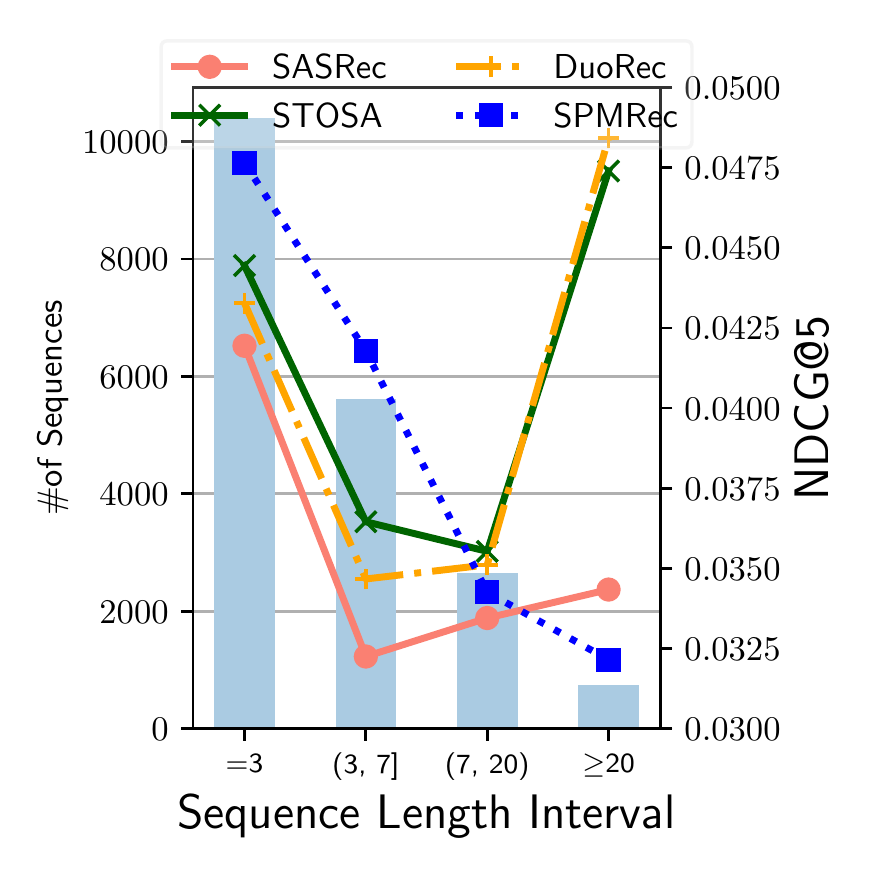}
    \caption{Toys}
    \label{fig:toys_item}
\end{subfigure}%\\
\begin{subfigure}[t]{0.23\textwidth}
    % \
    \includegraphics[width=\textwidth]{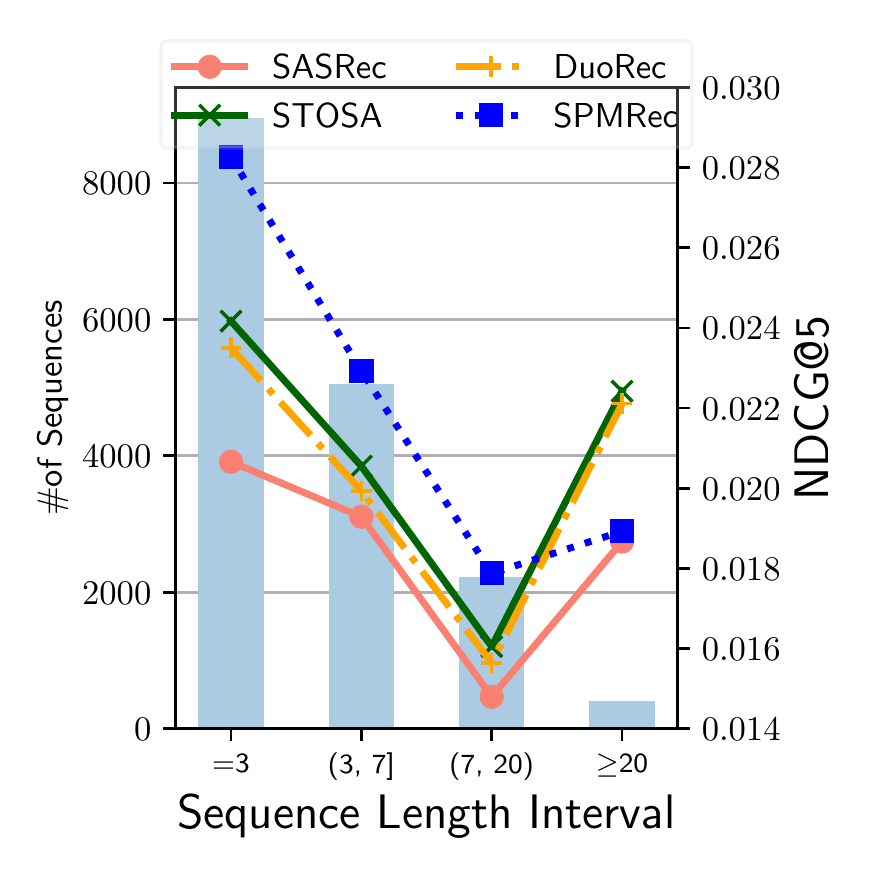}
    \caption{Tools}
    \label{fig:tools_item}
\end{subfigure}
\begin{subfigure}[t]{.23\textwidth}
    % \centering
    \includegraphics[width=\textwidth]{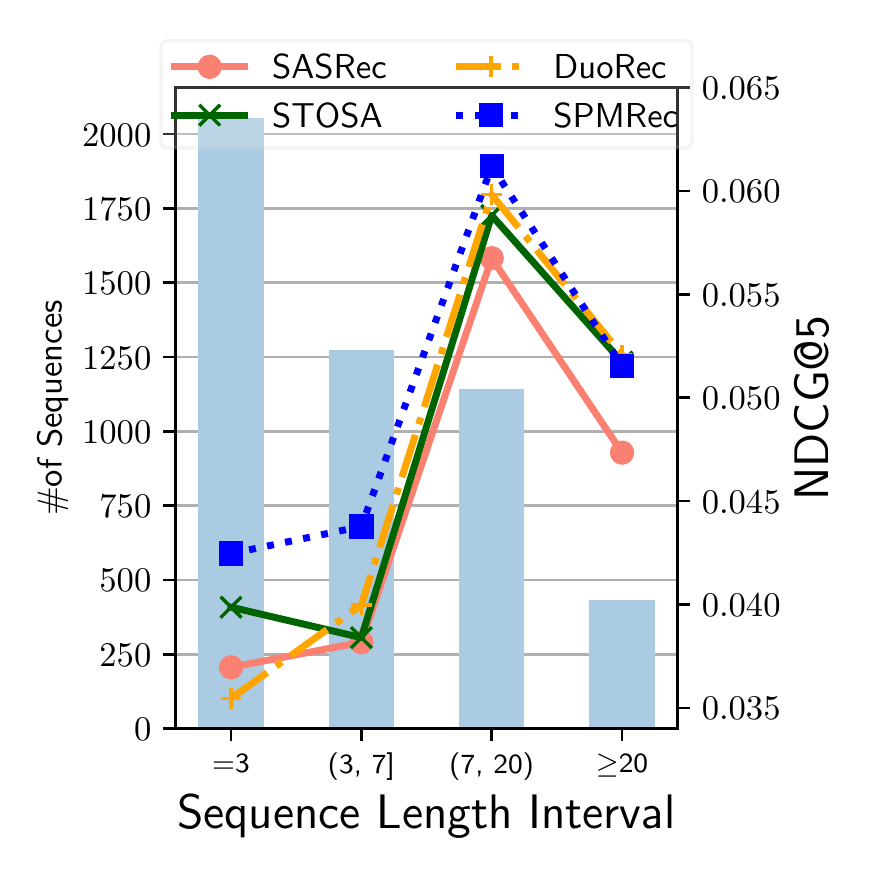}
    \caption{Office}
    \label{fig:office_item}
\end{subfigure}
% \vspace{-3mm}
\caption{NDCG@5 on different sequence groups based on sequence length.}
\label{fig:ndcg5_seqlen}
\end{figure*}

\textbf{\modelname is highly effective in short sequences recommendation.} 
We group sequences~(users) based on the number of training interactions~(sequence length) of users. Within each group, we average the testing performances of sequences in the group. The comparisons on each group are shown in Fig.~(\ref{fig:ndcg5_seqlen}). We can observe that \modelname achieves significant improvements in short sequences. Specifically, the improvements over the best STOSA range from 7.21\% to 23.04\%. The recommendation for short sequences is challenging due to limited interaction signals, which is known as the cold start problem. \modelname regularizes the sequence output embeddings and prevents the degeneration so that short sequences can be distinguishable and are properly modeled. We can also observe that \modelname has inferior performances in the longest sequence group, which is even worse than SASRec in the Toys dataset. However, the short sequences are in majority in both the used datasets and practical applications. With the proposed singular spectrum smoothing, \modelname benefits short sequences with the sacrifices of long sequences.

\begin{figure*}
\begin{subfigure}[t]{0.23\textwidth}
    % \
    \includegraphics[width=\textwidth]{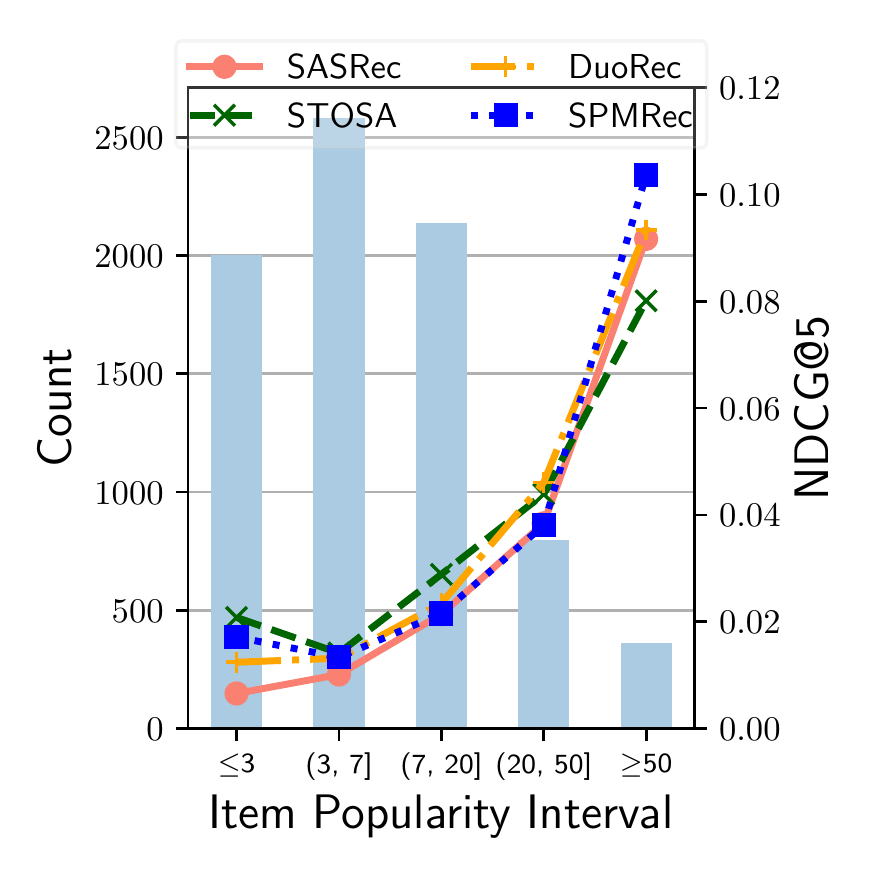}
    \caption{Beauty}
    \label{fig:beauty_item}
\end{subfigure}
\begin{subfigure}[t]{.23\textwidth}
    % \centering
    \includegraphics[width=\textwidth]{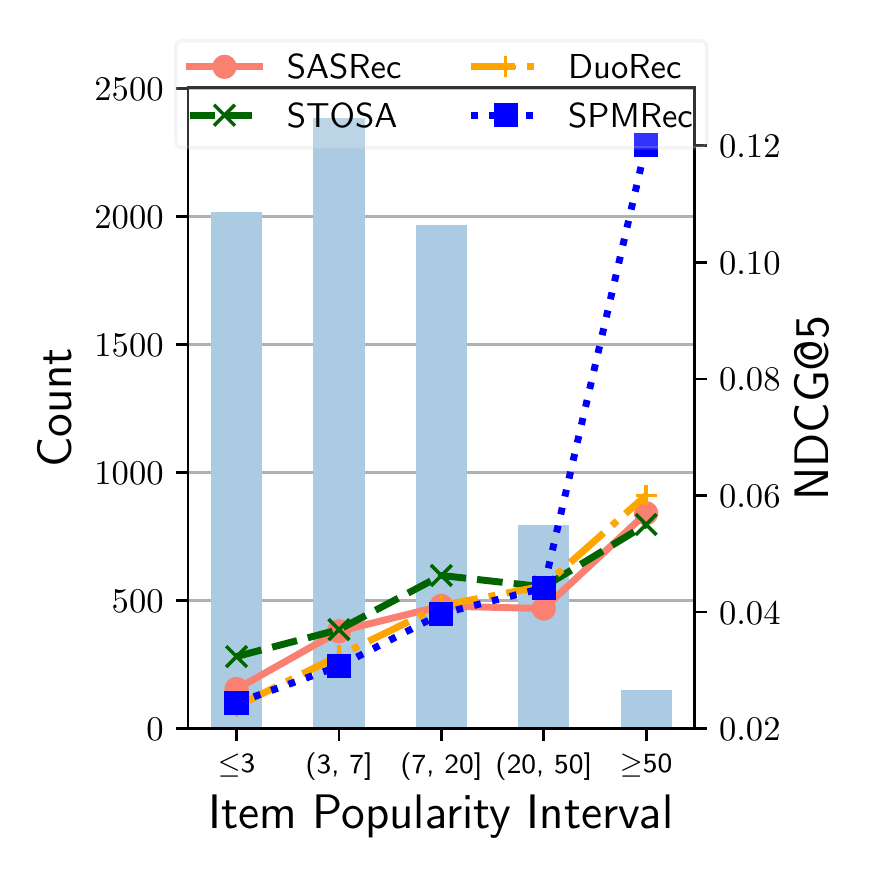}
    \caption{Toys}
    \label{fig:toys_item}
\end{subfigure}
% \vspace{-3mm}\\
% \\
\begin{subfigure}[t]{0.23\textwidth}
    % \
    \includegraphics[width=\textwidth]{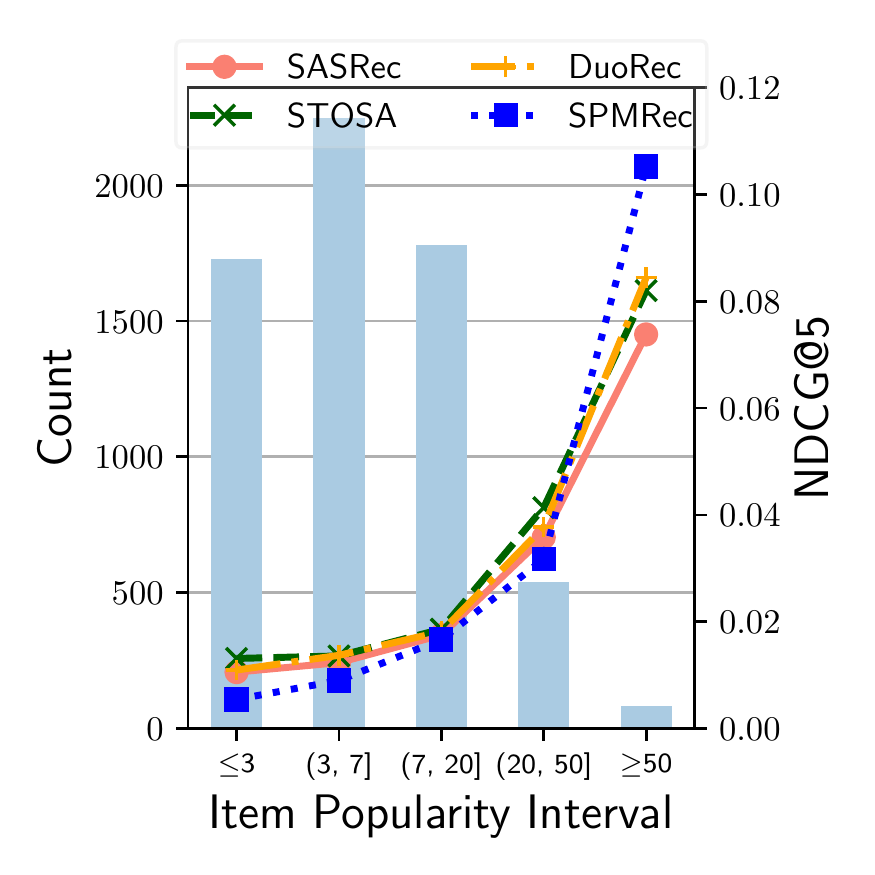}
    \caption{Tools}
    \label{fig:tools_item}
\end{subfigure}
\begin{subfigure}[t]{.23\textwidth}
    % \centering
    \includegraphics[width=\textwidth]{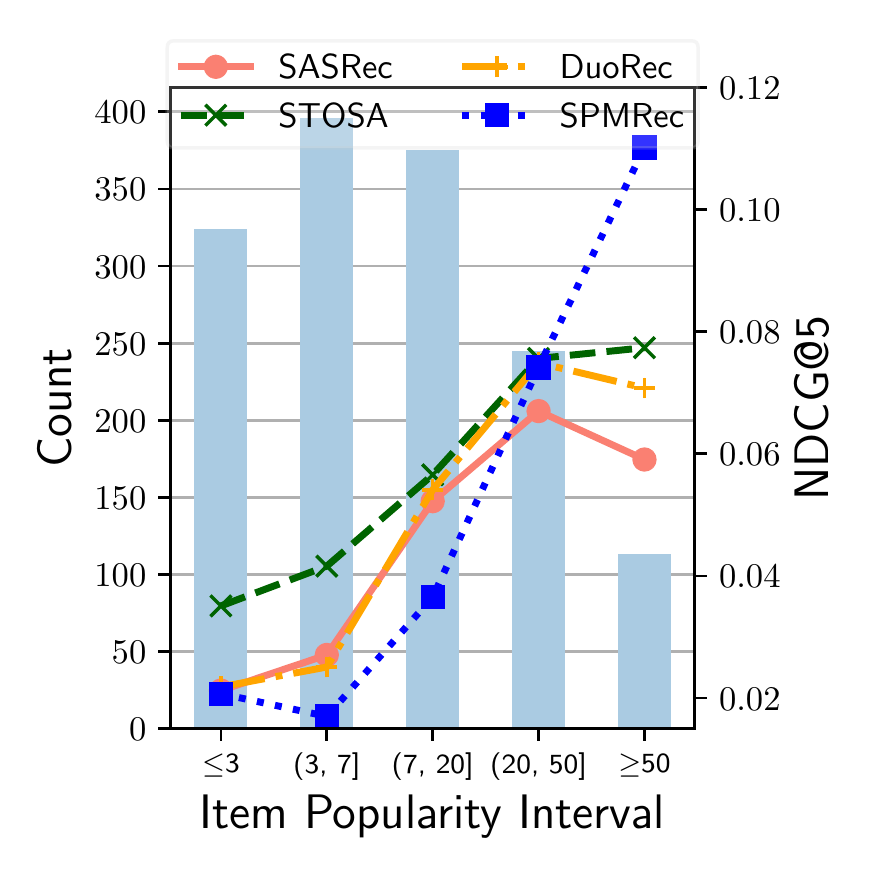}
    \caption{Office}
    \label{fig:office_item}
\end{subfigure}
% \vspace{-3mm}
\caption{NDCG@5 on different item groups based on item popularity.}
\label{fig:ndcg5_item}
\end{figure*}

\begin{figure}
\begin{subfigure}[t]{0.23\textwidth}
    % \
    \includegraphics[width=\textwidth]{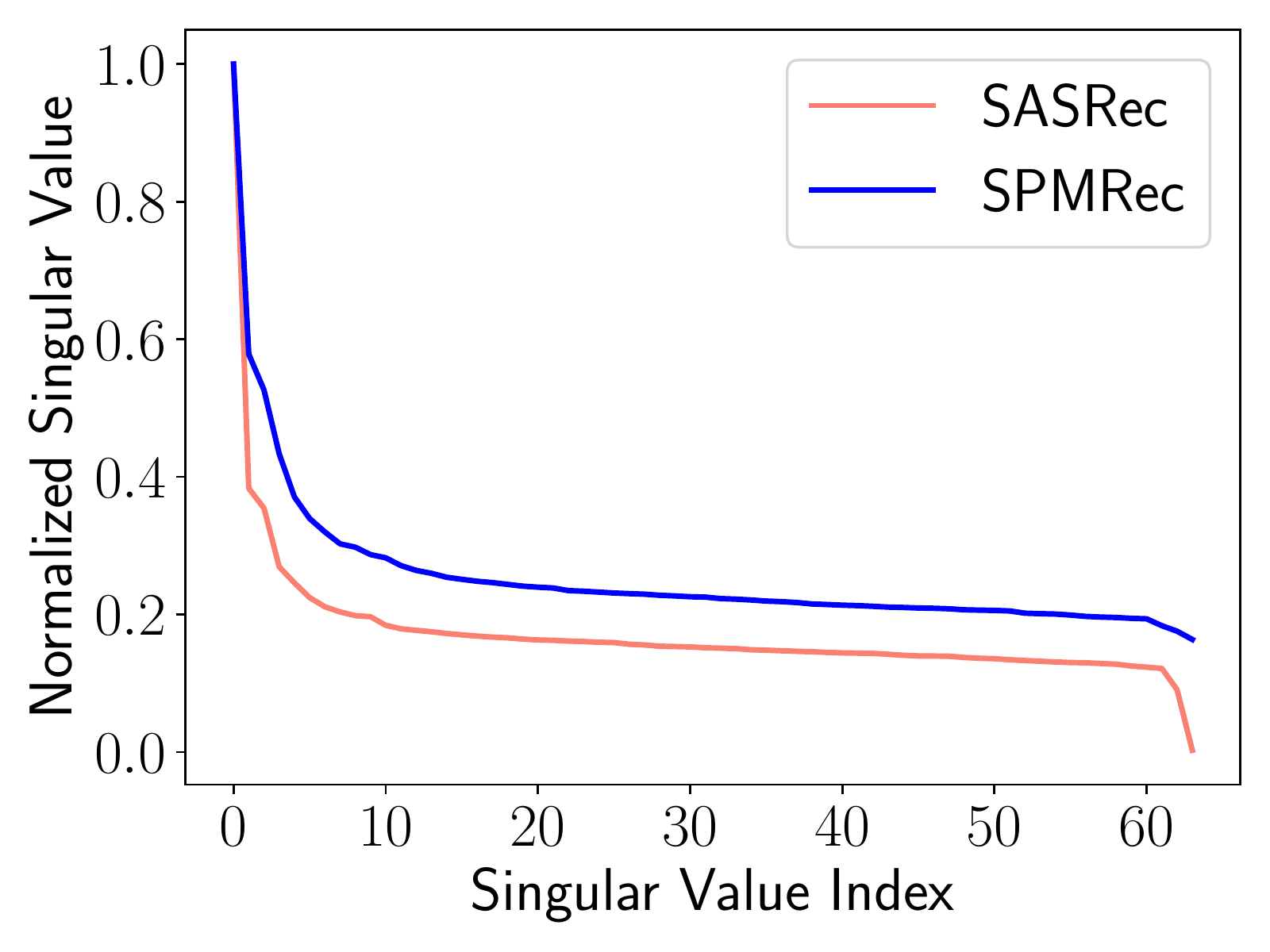}
    \caption{Beauty}
    \label{fig:beauty_item}
\end{subfigure}
\begin{subfigure}[t]{.23\textwidth}
    % \centering
    \includegraphics[width=\textwidth]{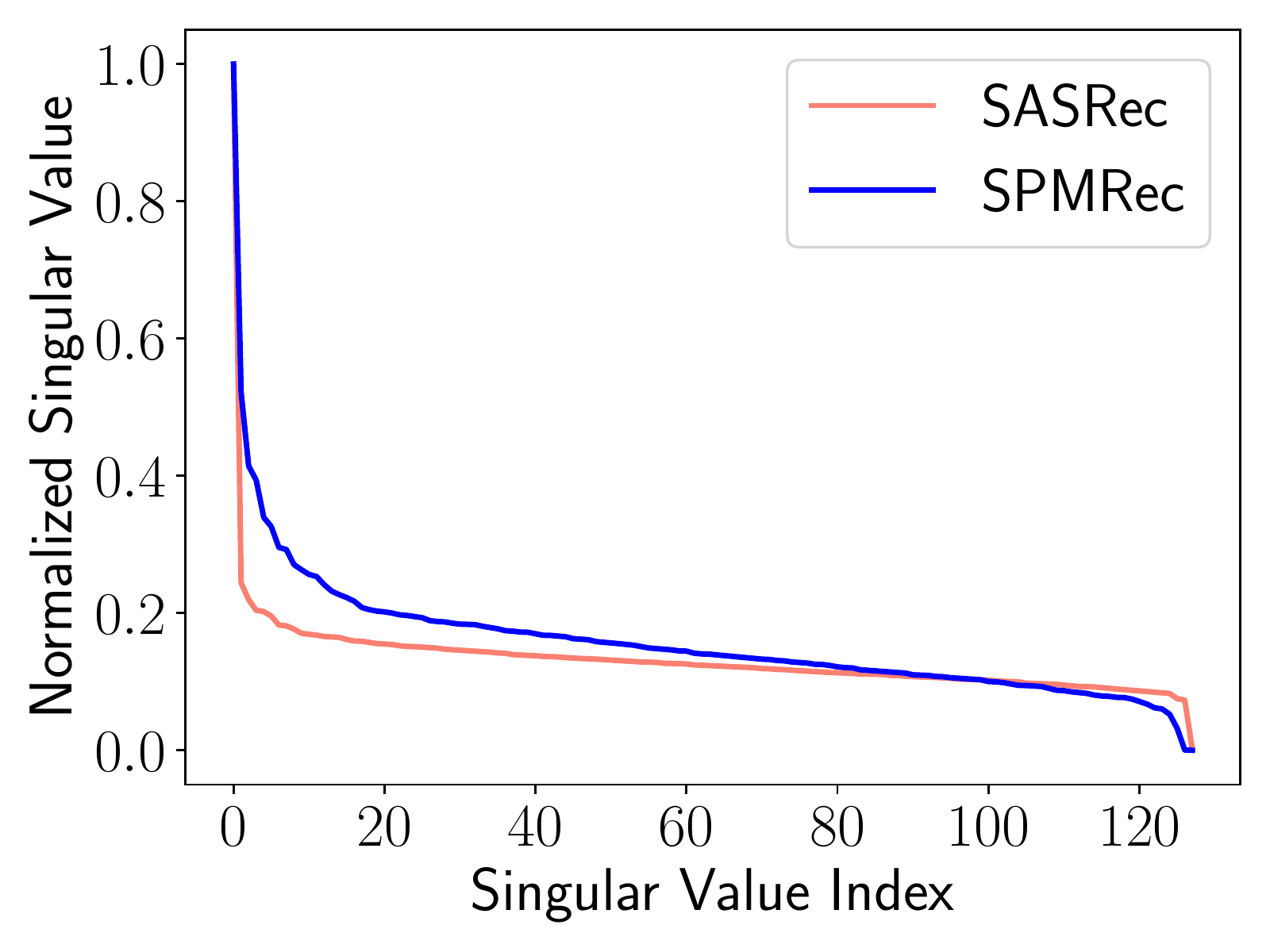}
    \caption{Toys}
    \label{fig:toys_item}
\end{subfigure}
% \vspace{-3mm}\\
\\
\begin{subfigure}[t]{0.23\textwidth}
    % \
    \includegraphics[width=\textwidth]{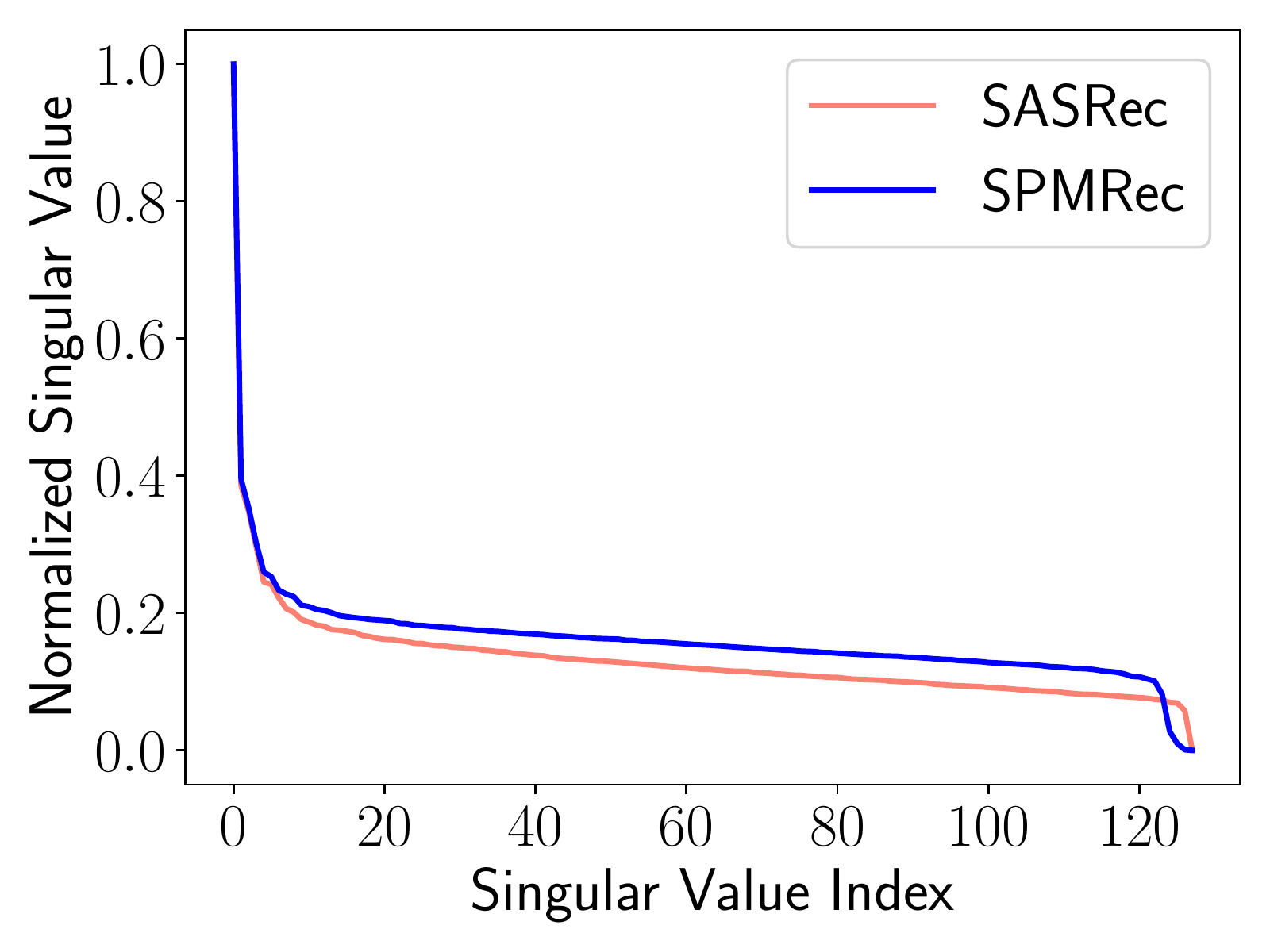}
    \caption{Tools}
    \label{fig:tools_item}
\end{subfigure}
\begin{subfigure}[t]{.23\textwidth}
    % \centering
    \includegraphics[width=\textwidth]{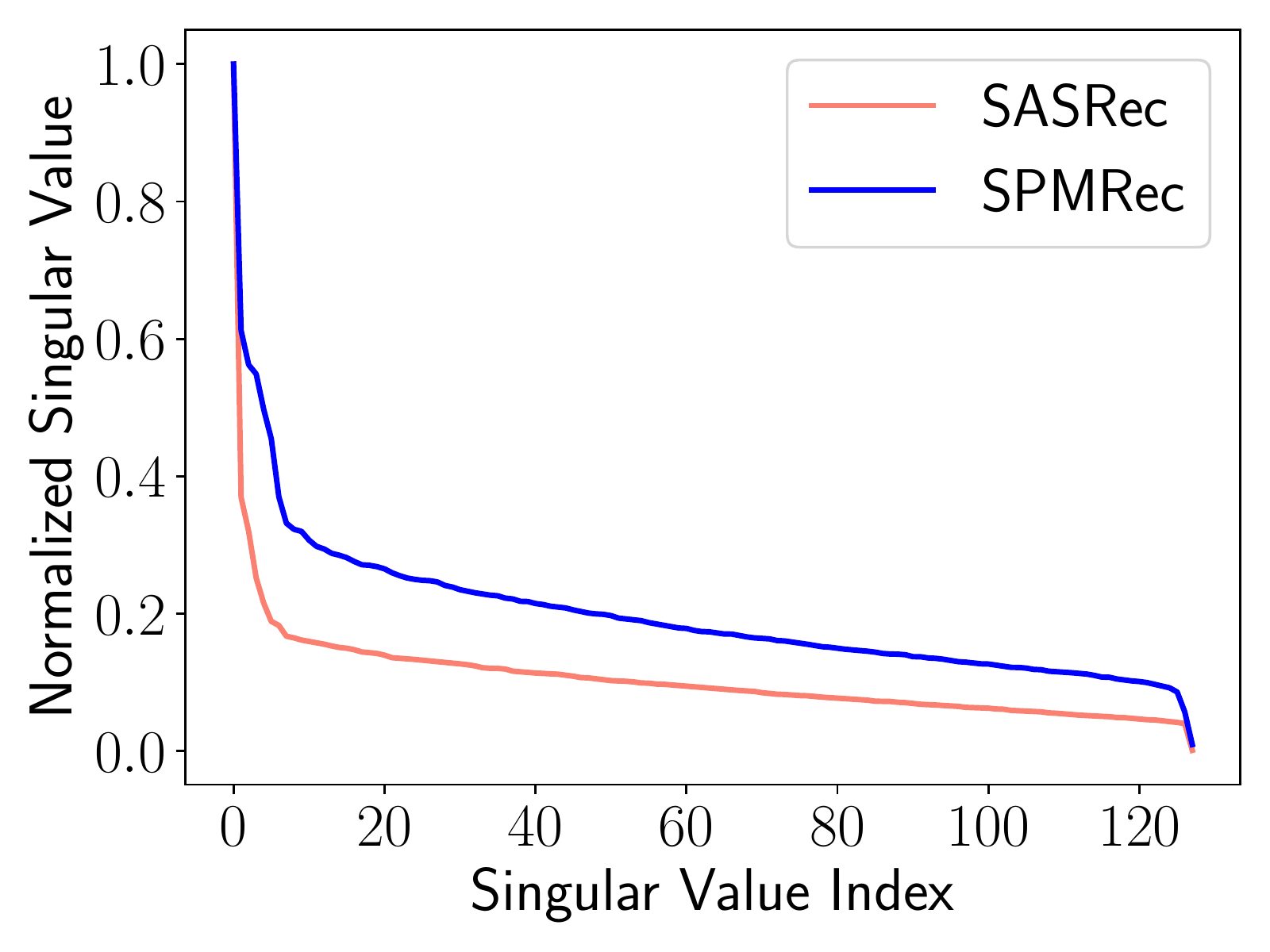}
    \caption{Office}
    \label{fig:office_item}
\end{subfigure}
% \vspace{-3mm}
\caption{Normalized Singular Values Curve Comparison between SASRec and the proposed \modelname in best grid searched settings.}
\label{fig:seq_singval_trend}
\end{figure}
\subsection{Sequential User Degeneration Analysis~(RQ5)}
We visualize the singular values curve of sequence output embedding for our proposed \modelname and SASRec, as shown in Fig.~(\ref{fig:seq_singval_trend}). We can observe that the proposed \modelname has a smoother decaying curve than SASRec. Moreover, for \modelname, the degrees of spectrum smoothness for all four datasets are 26.34 for Beauty, 20.76 for Tools, 20.25 for Toys, and 19.24 for Office. However, SASRec has lower degrees for spectrum smoothness as 19.54 for Beauty, 17.10 for Tools, 17.00 for Toys, and 13.67 for Office. \modelname consistently has larger degrees of spectrum smoothness than SASRec, which demonstrates the effectiveness of \modelname in preventing sequences from the degeneration for the sequential recommendation.

\section{Related Work}
% In this section, we discuss existing solutions in two related areas to our problem, including the sequential recommendation and representation degeneration problems. 

\subsection{Sequential Recommendation}
Sequential Recommendation (SR) is a technique that encodes the temporal interactions of each user as a sequence and uses the encoded behaviors to infer their preferences. The main idea behind SR is to model the transitions between items within sequences. The earliest SR methods used Markovian approaches, such as FPMC~\cite{rendle2010factorizing} and Fossil~\cite{he2016fusing}, with the assumption that the item prediction depends only on the previous interacted items
% , which is the first-order Markovian assumption. Fossil extends FPMC with the addition of item similarities information. 
Another line of work utilizes the Recurrent Neural Network~(RNN) as the base model for modeling sequential interactions. 
% The recurrent modeling characteristics of RNN make it suitable for modeling temporally ordered behaviors. 
Several works adopt RNNs to address the SR problem, including GRU4Rec~\cite{hidasi2015session}, HGN~\cite{ma2019hierarchical, peng2021ham}, HRNN~\cite{quadrana2017personalizing}, and RRN~\cite{wu2017recurrent}. Different variants of RNNs are proposed, such as the original GRU in GRU4Rec and hierarchical RNN in HGN and HRNN. 
The third line of methods to solve the SR problem is using the Convolution Neural Network~(CNN), with the interpretation of a sequence of item embeddings as an image, including Caser~\cite{tang2018personalized}, CosRec~\cite{yan2019cosrec}, and NextItNet~\cite{yuan2019simple}. 
% Caser proposes convolutions to latent dimensions and item perspectives for comprehensive sequence encoding. Instead of treating the sequence as an image, CosRec extends to represent the sequence as a 3D tensor and adopts the 2D convolution operators. The NextItNet proposes the dilated convolution in the sequence modeling. 
The recent success of the self-attention module~\cite{vaswani2017attention} inspires the adaptation of Transformer to the sequential recommendation task. The self-attention module captures the sequential high-order item transitions by modeling attention from all items in the sequence. SASRec~\cite{kang2018self} is the earliest work of applying Transformer to SR. BERT4Rec~\cite{sun2019bert4rec} builds upon BERT~\cite{devlin2018bert} with bidirectional Transformers for SR. FMLP-Rec~\cite{zhou2022filter} proposes an all-MLP architecture in the Transformer for SR. Several variants of Transformers~\cite{fan2021modeling, li2020time, wu2020sse, 10.1145/3404835.3463036} are further proposed.

\subsection{Representation Degeneration}
The representation degeneration problem describes the issue of learned embeddings being narrowly distributed in a small cone~\cite{DBLP:conf/iclr/GaoHTQWL19}. The degeneration issue is firstly proposed in the word representations learning problem. As shown in~\cite{DBLP:conf/iclr/GaoHTQWL19}, words from text data follow the long-tail distribution, which degenerates the representation learning process. 
Theoretically, it causes the narrow distribution of words in the representation latent space~\cite{DBLP:conf/iclr/GaoHTQWL19}. 
Moreover, an empirical analysis~\cite{DBLP:conf/acl/YuSK0RY22} shows that the learning of tail words degenerates the representation optimization of frequent words. An important analysis in~\cite{wang2020improving} shows that degeneration is closely related to the fast decaying singular values distribution. 
In other words, the largest singular value is significantly larger than all other singular values. 

Distinct from the natural language processing~(NLP) area, SR still has several unique characteristics. For example, sequences are short~(data sparsity in recommendation systems), while sentences\slash paragraphs in NLP are typically long. This distinct characteristic introduces additional challenges for SR, where the degeneration issue can also happen in the sequence output embeddings. 
DuoRec~\cite{qiu2022contrastive} observes that the representation degeneration issue also exists in the self-supervised SR, with data augmentations. DuoRec and DirectAU~\cite{DBLP:conf/kdd/WangYM000M22} optimizes the uniformity property in the in-batch representations to enforce the representations to be uniformly distributed, where the uniformity is based on the Euclidean distance. 
This work connects the item embeddings degeneration with the degeneration of sequence representations. The intertwining of both sequence and item rank degeneration is rarely discussed before in SR.
Moreover, some solutions from NLP to the degeneration problem only apply to the item side, including the CosReg~\cite{DBLP:conf/iclr/GaoHTQWL19} based on cosine similarities, Euclidean regularization based on learned word similarities~\cite{DBLP:conf/emnlp/0004GXMYS20}, spectrum control~\cite{wang2020improving, yan2022addressing}, and gradient gating~\cite{DBLP:conf/acl/YuSK0RY22}. 
% The general idea of these works is regularizing the representation matrix by strengthening the differences between representations, mostly either based on cosine similarities~(\textit{e.g.,} CosReg~\cite{DBLP:conf/iclr/GaoHTQWL19}) or Euclidean methods~(\textit{e.g.,} \cite{DBLP:conf/emnlp/0004GXMYS20} and the uniformity property adopted by DuoRec and DirectAU). Unlike existing solutions, this work investigates the rank degeneration issue from the singular value perspective of representations and proposes a novel regularization, which preserves the rank of representations and keeps representations distinguishable.
% The existing solutions to this degeneration problem include the CosReg~\cite{DBLP:conf/iclr/GaoHTQWL19} based on cosine similarities, Euclidean regularization based on learned word similarities~\cite{DBLP:conf/emnlp/0004GXMYS20}, spectrum control~\cite{wang2020improving}, and gradient gating~\cite{DBLP:conf/acl/YuSK0RY22}. The general idea of these works is either regularizing the representation matrix directly~\cite{DBLP:conf/iclr/GaoHTQWL19,wang2020improving,DBLP:conf/emnlp/0004GXMYS20} or modifying the gradient component of word embeddings optimization~\cite{DBLP:conf/acl/YuSK0RY22}.

As discussed in~\cite{de2021transformers4rec}, natural language processing~(NLP) and sequential recommendation~(SR) share multiple similarities, including both learning discrete tokens learning~(words in NLP and items in SR) and long-tail distribution of frequency for both words and items. With this inspiration, DuoRec~\cite{qiu2022contrastive} observes that the representation degeneration issue also exists in the self-supervised SR, with data augmentations. DuoRec optimizes the uniformity property in the in-batch augmented representations to alleviate the degeneration issue in contrastive learning, where the uniformity is based on the Euclidean distance.

\section{Conclusion}
We theoretically and empirically identify the relationship between the representation degeneration and recommendation diversity. Motivated by this connection, we propose a novel metric to measure the degree of degeneration as the area under the singular value curve~(AUSC) to constrain the sequence and item embeddings learning. We further propose a singular spectrum smoothing regularization as a surrogate of AUSC to alleviate the degeneration and improve the diversity simultaneously. We empirically demonstrate the effectiveness of the proposed singular spectrum smoothing regularization in recommendation performances, especially for short sequences and popular items. We also empirically demonstrate a strong connection between the proposed singular spectrum smoothing regularization and the recommendation diversity.
\bibliographystyle{ACM-Reference-Format}
\bibliography{sample-base}

%%
%% If your work has an appendix, this is the place to put it.
\appendix

\section{Baselines and Hyper-parameters Grid Search}
We implement \modelname with Pytorch based on the S3-Rec~\cite{zhou2020s3} code base. We grid search all parameters and report the test performance based on the best validation results.
We search the representation dimension in $\{64, 128\}$, maximum sequence length from $\{50, 100\}$, learning rate in $\{10^{-3},10^{-4}\}$, the $L2$ regularization control weight from $\{0.0, 10^{-1}, 10^{-2}, 10^{-3}\}$, dropout rate from $\{0.3, 0.5, 0.7\}$ for all methods. 
For sequential methods, we search number of layers from $\{1,2,3\}$, and number of heads in $\{1,2,4\}$. 
We adopt the early stopping strategy that model optimization stops when the validation MRR does not increase for 50 epochs. 
The followings are the model specific hyper-parameters search ranges of baselines:
\begin{itemize}[leftmargin=*]
    \item \textbf{BPR\footnote{\url{https://github.com/xiangwang1223/neural_graph_collaborative_filtering}}:} BPR is the most classical collaborative filtering method for personalized ranking with implicit feedbacks. We search the learning rate in $\{10^{-3},10^{-4}\}$ and $L2$ regularization weight from $\{10^{-1}, 10^{-2}, 10^{-3}\}$.
    \item \textbf{Caser\footnote{\url{https://github.com/graytowne/caser_pytorch}}:} A CNN-based sequential recommendation method that views the sequence embedding matrix as an image and applies convolution operators to it. We search the length $L$ from $\{5, 10\}$, and $T$ from $\{1, 3, 5\}$.
    \item \textbf{SASRec\footnote{\url{https://github.com/RUCAIBox/CIKM2020-S3Rec}}:} The state-of-the-art sequential method that depends on the Transformer architecture. We search the dropout rate from $\{0.3, 0.5, 0.7\}$. 
    \item \textbf{BERT4Rec\footnote{\url{https://github.com/FeiSun/BERT4Rec}}:} This method extends SASRec to model bidirectional item transitions with standard Cloze objective. We search the mask probability from the range of $\{0.1, 0.2, 0.3, 0.5, 0.7\}$.
    \item \textbf{DT4SR\footnote{\url{https://github.com/DyGRec/DT4SR}}:} A metric learning-base sequential method that models items as distributions and proposes mean and covariance Transformers. We search the dropout rate from $\{0.3, 0.5, 0.7\}$.
    \item \textbf{STOSA\footnote{\url{https://github.com/zfan20/STOSA}}:} A metric learning-base sequential method that models items as distributions and proposes a Wasserstein self-attention module. We search the dropout rate from $\{0.3, 0.5, 0.7\}$.
    \item \textbf{DuoRec:}\footnote{\url{https://github.com/RuihongQiu/DuoRec}} This method introduces unsupervised Dropout and supervised semantic augmentations in self-supervised learning for sequential recommendation. 
    \item \textbf{FMLP-Rec:}\footnote{\url{https://github.com/Woeee/FMLP-Rec}} Another most recent SR method adopting the full mlp layers architecture and filter methods on the self-attention module. 
\end{itemize}

\end{document}